\documentclass[useAMS,usenatbib]{mn2e}
\usepackage[english]{babel}
\usepackage[latin1]{inputenc}
\usepackage{graphicx}
\usepackage{longtable}
\title[]{Eclipsing binary stars in the Large Magellanic Cloud. Results from the EROS-2, OGLE and VMC surveys\thanks{Based on observations made with VISTA at the Paranal Observatory under program ID
179.B-2003.}}
\author[T. Muraveva et al.]{T. Muraveva$^{1}$, G. Clementini$^{1}$, C. Maceroni$^{2}$, C. J. Evans$^{3}$, M. I. Moretti$^{1,4,5}$,  
\newauthor M.-R.L. Cioni$^{6,7}$, J. B. Marquette$^{8}$, V. Ripepi$^{4}$, R. de Grijs$^{9,10}$, 
\newauthor M. A. T. Groenewegen$^{11}$, A. E. Piatti$^{12,13}$, J. Th. van Loon$^{14}$\\
$^{1}$ INAF-Osservatorio Astronomico di Bologna, via Ranzani 1, Bologna, 40127,  Italy\\
$^{2}$ INAF-Osservatorio Astronomico di Roma, via di Frascati 33, Monteporzio Catone, 00040, Italy\\
$^{3}$ UK Astronomy Technology Centre, Royal Observatory Edinburgh, Blackford Hill, Edinburgh, EH9 3HJ, United Kingdom\\
$^{4}$ INAF-Osservatorio Astronomico di Capodimonte, via Moiariello
 16, Naples, 80131, Italy \\
$^{5}$ Scuola Normale Superiore di Pisa, piazza dei Cavalieri 7, Pisa, 56126, Italy\\
$^{6}$ University of Hertfordshire, Physics Astronomy and
 Mathematics, College Lane, Hatfield, AL10 9AB, United Kingdom\\
$^{7}$ Leibniz-Institut f\"{u}r Astrophysik Potsdam, An der Sternwarte 16, Potsdam, 14482, Germany\\ 
$^{8}$ UPMC-CNRS, UMR7095, Institut d'Astrophysique de Paris, F-75014, Paris, France\\
$^{9}$ Kavli Institute for Astronomy and Astrophysics, Peking University, Yi He Yuan Lu 5, Hai Dian District, Beijing, 100871, China\\
$^{10}$ Department of Astronomy, Peking University, Yi He Yuan Lu 5, Hai Dian District, Beijing, 100871, China\\
$^{11}$ Royal Observatory of Belgium, Ringlaan 3, B-1180 Brussels, Belgium\\  
$^{12}$ Observatorio Astron{\'o}mico, Universidad Nacional de C{\'o}rdoba, Laprida  854, 5000, C{\'o}rdoba, Argentina\\
$^{13}$ Consejo Nacional de Investigaciones Cient\'{\i}ficas y T{\'e}cnicas, Av. Rivadavia 1917, C1033AAJ, Buenos Aires, Argentina\\
$^{14}$ Lennard-Jones Laboratories, Keele University, ST5 5BG, United Kingdom
}

\begin{document}

\date{Accepted 2014 June 6.  Received 2014 June 6; in original form 2014 April 14}

\pagerange{\pageref{firstpage}--\pageref{lastpage}} \pubyear{2002}

\maketitle

\label{firstpage}

\begin{abstract}
We present a catalogue of 1768 eclipsing binary stars (EBs) detected in the Large Magellanic Cloud (LMC) by the second generation of the EROS survey (hereinafter EROS-2); 493 of them are new discoveries located in outer regions (out of the central bar) %Richard
of the LMC. %not yet observed by other surveys. 
These sources were originally included in a list of  candidate  classical Cepheids (CCs) extracted from the EROS-2 catalogue on the basis of the period (0.89 $<P_{EROS}<$15.85 days) versus luminosity ($13.39 < \langle B_{EROS}\rangle<17.82$ mag) diagram.  After visual inspection of the light curves we reclassified them  as eclipsing binaries.
They have blue colours ($B_{EROS} - R_{EROS}  < $ 0.2 mag)  hence  we classed them as hot eclipsing binaries (HEBs) containing hot massive components: main sequence (MS) stars or blue giants. %, which we classed as hot eclipsing binaries (HEBs). 
 We present  $K_{\rm s}$-band  light curves for 999 binaries from  our sample that have a counterpart in the VISTA near-infrared ESO public survey of the Magellanic Clouds system (VMC). We provide spectral classifications of 13 HEBs with existing spectroscopy. We divided our sample into contact-like  binaries and detached/semi-detached systems based on both  visual inspection and the parameters of the Fourier decomposition of the light curves %. We derived  $K_{\rm s}$ band magnitudes at maximum light for  999 binaries which have a counterpart in the VISTA for Magellanic Cloud  (VMC), near-infrared ESO public survey. 
and analysed the period-luminosity  ($PL$) relations  of the contact-like systems using the $R_{EROS}$ and $K_{\rm s}$ magnitudes at maximum light.  
The contact-like binaries in our sample do not  follow $PL$  relations. %in the $K_{\rm s}$-band  ($PL_K$) 
We analysed the sample of contact binaries from the OGLE~III catalogue and  confirmed that  
 $PL_I$  and $PL_{K_{\rm s}}$ sequences are defined only by  eclipsing binaries containing a red giant component.%Maria-Rosa comment
%The HEBs presented in this paper mainly trace regions of most recent star formation activity in the LMC and could  thus be used to study the galaxy internal structure.   

\end{abstract}
Stars: binaries: eclipsing -- galaxies: Magellanic Clouds -- surveys -- techniques: photometric
\begin{keywords}

\end{keywords}

\section{Introduction}

%Tania: I rewrote this part according to comments of Martin and Ida. They asked to add more information about classification of binaries and about  W UMa type.
Eclipsing binary stars (EBs) are particularly important to study stellar evolution and stellar structure, since some fundamental stellar parameters can be determined using geometrical constraints of the systems. Masses are estimated dynamically via radial velocities, radii from the eclipse durations and the temperature ratio {(strictly the surface-brightness ratio) from the eclipse depths. The  radii and temperature together are used to measure the luminosity of the system. From the estimated luminosity and observed fluxes the distances to the EBs can be determined. This method requires photometric and spectroscopic data (see reviews by \citealt{And1991}; \citealt{Tor2010}) and is used to measure the distances to nearby galaxies.  Recently, \citet{Pietr2013} used EBs to measure a distance to the Large Magellanic Cloud (LMC) which is considered to be accurate to 2\%. Moreover, analysis of EBs  allows to estimate stellar ages by comparing the mass-radius relation with stellar evolution models. %information from LSST Science Book, suggested by Maria-Rosa

The classification of EBs is based on the distance between components, relative to their sizes.  If the two components do not fill their Roche lobes, the system is considered to be a detached binary. In a semi-detached binary one of the two components fills its Roche lobe and mass transfer occurs.  In  contact eclipsing binaries both stars fill their Roche lobes. To classify contact and detached/semi-detached binaries  analysis of the Fourier parameters (see Section~\ref{fourier}) and visual inspection of the light curves are necessary. W UMa type stars are contact binary systems with orbital periods typically less than 1 day, composed of  main sequence (MS) turn-off stars \citep{Ruc1998}. It was shown that this class of binaries could be used as distance indicators  since the size of  the two components  could be determined from the orbital period, which in combination with the colour information allows one to derive absolute magnitudes (\citealt{Ruc1997}, and references therein).  Indeed, \citet{Ruc1997} used the method  to determine the distance to W UMa-type contact systems in the Galactic Bulge. 

\citet{Ruc1998} analysed contact systems with orbital periods longer than 1 day, discovered in Baade's Window  by the OGLE project. He found that there are two classes of such EBs that had passed the Fourier light-curve shape filter. The first group included W UMa type systems with orbital periods extending to $P<1.3-1.5$ days. These objects could be the most massive representatives of the population of old, close binary systems that are entering the final stages of evolution, before merging of the two components, or analogues of the contact blue stragglers in globular clusters.  The second group contained long-period red binaries.  \citet{Ruc1998} suggested that the latter are not contact binaries, but rather systems composed by a red giant or subgiant, close to filling its Roche lobe, that is tidally distorted by the interaction with the companion which is assumed to be invisible (a MS star or a collapsed object).  The source of variability is therefore an  ellipsoidal variation of the distorted star. 
%According to the fact that Fourier filter passes genuine contact binaries as well as ellipsoidal variables that have contact-binary like shape of light curve, we call all objects, passed through the Fourier filter and the visual inspection of the light curve, contact-like systems. 
Given that the Fourier filter passes genuine contact binaries as well as sources which have light curves with a contact-binary-like shape such as the ellipsoidal variables, we classify all systems which satisfy the Fourier analysis/visual inspection as contact-like binaries.
While W UMa type stars in the Galaxy seem to be limited to  periods $P<1.3-1.5$ days, massive, young,  blue systems of W UMa type with longer orbital periods of 2-3 days exist in the LMC \citep{Ruc1999}. These objects may follow a period-luminosity ($PL$) relation as suggested by \citet{Ruc1999}. One of the aims of the current research was to check this statement.%I added the last phrase because Andres asked why are we checking the possibility to measure the distance if we stated it on the first page.

  In the past years a number of microlensing surveys such as OGLE \citep{Ud1997}, MACHO \citep{Al1997} and EROS \citep{Tiss2007} have discovered thousands of variable stars of different types in the Magellanic Clouds. A significant fraction of these variables are EBs. Nine different catalogues of EBs detected in the LMC by the microlensing surveys have been presented. During the first stage of the EROS survey, 79 candidate EBs were identified in the bar of the LMC \citep{Gris1995}. The MACHO survey identified an initial sample of 611 LMC EBs \citep{Al1997}. Subsequently, \citet{Der2007} reanalysed the  eclipsing variables in the MACHO database, corrected their periods  and presented a ``clean" sample of 3031 EBs. \citet{Facc2007} provided a new sample of 4634 EBs in the LMC from the MACHO catalogue, expanding the previous sample of 611 objects from \citet{Al1997}. 
%The cross-correlation with  \citet{Der2007} and \citet{Facc2007} catalogues of LMC EBs shows that 797 objects were already known 
%(grey dots in panel (b) of Fig.~\ref{cat}, in red in the electronic edition of the journal).  
%\citet{Al1997} presented a catalogue of 611 EBs  detected by the MACHO survey, \citet{Facc2007} expanded this catalogue to 4634 stars. \citet{Der2007} reanalysed light curves of eclipsing %binaries from the MACHO database and presented a sample of 3031 EBs.  
Using the OGLE~II data 3332 EBs were identified in the LMC (\citealt{Wyr2003}, \citealt{Groen2005}, \citealt{Grac2010}).  \citet{Grac2011} provided a sample of 26121 LMC EBs  detected by the OGLE~III survey.  Finally, \citet{Sos12} identified 1377 EBs and 156 ellipsoidal variables in the Gaia South Ecliptic Pole (GSEP\footnote{The GSEP is an area of about
5.3 deg$^2$ around the South Ecliptic Pole, of which a central rhombus-shaped portion of  5 $\times$ 0.7 deg$^2$ corresponds to the region that the European Space Agency satellite Gaia, launched on December 19th,  2013,  is observing repeatedly during commissioning.}) area based on the OGLE~IV survey. 

In this paper we present a catalogue of 1768 EBs observed in the LMC by the EROS-2 survey. This survey monitored both the inner and outer regions of the LMC and more than a quarter of the EBs in our sample are new discoveries,  not yet identified  by other surveys. 
The stars were originally selected as candidate  classical Cepheids (CCs). % on the basis of the period (0.89 $<P_{EROS}<$15.85 d) versus luminosity ($13.39 < B_{EROS}<17.82$ mag) diagram (see \citealt{Mor2014}).  
However,  visual inspection of their light curves and the position on the colour-magnitude diagram (CMD) revealed they are, in fact, EBs containing hot massive components (MS stars or blue giants).  %For this reason we named them HEBs.
%
%The EROS-2 survey has provided so far the largest spatial coverage of the LMC. A large fraction of the EBs discussed in the paper  are mainly  concentrated near the bar
% 493 of them are new discoveries located in outer regions of the LMC not yet observed  by any other surveys.   EROS-2  has provided so far the largest spatial coverage of the LMC and 
These bright, relatively young objects (mainly  concentrated near the central bar and northwest spiral arm) trace the regions of recent star-formation activity in the LMC.  We present the results of the analysis of  the 1768 EBs and discuss the general properties of the sample. 
 
 %I removed this part because of the Martin's comment.
%One aim of the current research was to check  whether we can use %the possibility of using the contact systems in our sample as distance indicators.
We apply \citet{Ruc1997}'s method to classify the binaries by Fourier analysis of the light curves  and  divide our sample  into contact-binary-like and non-contact (detached and semi-detached)  systems. We then check whether the contact-binary-like systems follow a $PL$ relation using the $R_{EROS}$ magnitudes and the  %in the $K_{\rm s}$-band  ($PL_K$) using 
$K_{\rm s}$-band data provided by the VISTA near-infrared $YJK_{\rm s}$
survey of the Magellanic Clouds system \citep[VMC;][]{Cioni2011}. Using the VMC $K_{\rm s}$-data we also investigate the near-infrared  $PL$  relation of the whole sample of contact binaries in the OGLE~III catalogue.

In Section~2 we describe the method of selection and the general properties of our EROS-2 sample of EBs. In Section~3 we outline the procedure we used to classify the contact binaries. In Section~4 we present the $PL$ relations %in the $K_{\rm s}$-band 
of contact binaries extracted from the EROS-2 and OGLE~III catalogues. 
%In Section~5 we briefly compare the structure of the LMC traced by the CCs and by the hot EBs. 
A summary of our results is provided in Section 5.

\section{Data}

\subsection{EROS-2 sample}
The EROS-2  microlensing survey has monitored about 88 deg$^2$ of the LMC discovering a large number of CCs, RR Lyrae stars, binaries and long period variables (LPVs), both in the centre and in the outer regions of the galaxy. The survey was carried out with the Marly 1-m telescope at ESO, La Silla, from July 1996 to February 2003.  Observations were performed in two wide passbands, the so-called $R_{EROS}$ band centered close to the $I_C$ standard band, and the $B_{EROS}$ band intermediate between the standard $V$ and $R$ bands\footnote{The $B_{EROS}$ passband covers the wavelength interval from 420 to 720 nm, %that overlaps to the $V$ and $R$ standard bands, and a red channel with ? = (
the $R_{EROS}$ passband covers the interval from  620  to 920 nm.} \citep{Tiss2007}. The EROS magnitudes can be transformed to the Johnson-Cousins  standard system to  a   precision of $\sim$0.1\ mag, using the following equations from \citet{Tiss2007}:
\begin{equation}\label{I-band}
R_{EROS} = I_C   
\end{equation}
\begin{equation}\label{V-band}
B_{EROS}=V_J-0.4(V_J-I_C)
\end{equation}

%JB comment
The detection of variable stars and the determination of periods ($P_{EROS}$) were performed by an automatic pipeline based on the Analysis of Variance (AoV) method and software developed by \citet{Beaul1997} and \citet{Sc_Cz2003} (see \citealt{Marq2009} and references therein for further details). %From Ida's paper
As part of the collaboration between EROS-2 and VMC team members  
a sample of  5800 LMC candidate CCs was extracted from  the whole EROS-2 photometric catalogue of candidate variable stars by visually selecting them %specific sources were visually selected 
in the $B_{EROS}$ {\it versus} $P_{EROS}$ diagram on the basis of the $PL$ relation of CCs. The right panel of Figure 6 of \citet{Mor2014} shows how the selection was performed. The selected candidate CCs have EROS-2 periods  in the range $0.89<P_{EROS}<15.85$ days, and   mean $B_{EROS}$ magnitudes in the range  $13.39 < \langle B_{EROS}\rangle<17.82$ mag.  %These limits were visually set 
The faint magnitude/short period limit allows to reduce the contamination of the candidate CC sample by shorter period variables, such as the RR Lyrae stars, 
% at the faint magnitude/short period  side of the $PL$ relation, 
whereas  %reflect the bright limit of  the EROS-2 data  available to us at  
the bright magnitude/long period limit reflects the bright cut of  the EROS-2 data  available to us
%side of the  distribution
  (see Figure 6 of \citealt{Mor2014}).   %The EROS-2 collaboration kindly provided us  
%$B_{EROS}$ and $R_{EROS}$ single-epoch data
 $R_{EROS}$ and $B_{EROS}$  time-series photometry (and related errors),   periods,  and mean magnitudes for these 5800 LMC sources  were kindly made available to us by the EROS-2 team. 
%{\bf The catalogue of 176 EBs towards the Galactic spiral arms from the EROS-2 survey was provided by \citet{Der2002}, but there was not search for the EROS-2 eclipsing %%%binaries in the LMC so far.}
%classified as candidate CCs  because in the $P_{EROS}$ versus $B_{EROS}$ diagram they fall in the region  $13.39 < B_{EROS}<17.82$ mag, and $0.89<P_{EROS}<15.85$ days. 
%candidate CCs
%which were extracted from EROS-2 photometric catalogue . 
%
%The catalogue contains information about periods ($P_{EROS}$) and mean magnitudes of objects in the $B_{EROS}$ and $R_{EROS}$ passbands. Stars were classified as candidate CCs  %on the base of the $P_{EROS}$ versus $B_{EROS}$ diagram by consideration of objects with  $13.39 < B_{EROS}<17.82$ mag, and $0.89<P_{EROS}<15.85$ days. %Gisella, should I put %here the Figure with the selection of the EROS-2 - PL relation, similar to Ida's paper?
 Figure~\ref{cmd_eros} shows the $V$, $V-I$ CMD of the candidate CCs after transforming the $B_{EROS}$, $R_{EROS}$ average magnitudes of the sources to $V$, $I$ standard magnitudes by applying equations~\ref{I-band} and \ref{V-band}.
As it could be seen in Fig.~\ref{cmd_eros} the candidate CCs are distributed in three different regions of the CMD:  $(V-I)< 0.33$ mag (2085 objects), $0.33\leq(V-I)<1.67$ mag (3402 stars) and $(V-I)\geq1.67$ mag (313 objects). These regions are marked by dashed lines in the figure.
We visually inspected  the light curves of the stars in all three samples with the GRaphical Analyser of TIme Series (GRATIS) software developed at the Bologna Observatory by P. Montegriffo (see, e.g., \citealt{Cl2000}), and discovered that the majority of the objects with $V-I$ colours bluer than  0.33 mag are EBs; objects with colours $0.33\leq(V-I)<1.67$ mag are bona-fide CCs, and stars with colours $(V-I)\geq1.67$ mag are LPVs. The latter objects  accidentally fall in the sample of the candidate CCs, according to the $P_{EROS}$ and $B_{EROS}$ criteria,
% (Figure~\ref{er_pl})
because their EROS periods are aliases of the actual periods, which are usually in the range from tens to thousands of days.

The EROS-2 visual photometry of the bona-fide CCs is being used in combination with the VMC $K_{\rm s}$ photometry to study the $PL_{K_{\rm s}}$ relation of CCs and determine the distances to the different regions of the LMC \citep{Rip12a,Rip12b}. In the present paper we focus on the sample of EBs contaminating the candidate CCs. Hence, in the following sections we will specifically describe  the analysis of the 2085 sources with  $(V-I)< 0.33$ mag (this corresponds to objects with $B_{EROS} - R_{EROS} <$ 0.2 mag).  
%We explicitly note that our analysis is confined to the EBs which contaminate EROS-2 candidate CCs sample. 
We explicitly note that the EROS-2 catalogue of  candidate variables contains 
a much larger number of EBs, however, their study is beyond the purposes of the present paper.
\begin{figure}
 \includegraphics[width=8cm]{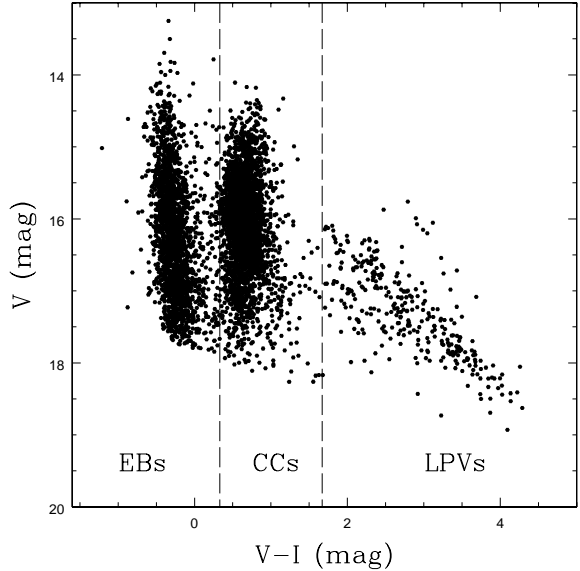}
   \caption{Colour-magnitude diagram of candidate CCs in the LMC, extracted from the EROS-2 catalogue on the basis of the period (0.89 $<P_{EROS}<$15.85 days) versus luminosity ($13.39 < B_{EROS}<17.82$ mag) diagram. The dashed lines correspond to $(V-I)=0.33$ mag and $(V-I)=1.67$ mag, respectively, and indicate the nature of the sources as classified in this study.}
  \label{cmd_eros}
\end{figure}
The analysis of the EBs contaminating the CC sample was performed running GRATIS on  the $R_{EROS}$ light curves  and showed that 83 objects from the sample are bona-fide CCs, 225 are small-amplitude variables, nine objects have light curves which are too noisy to be classified, and 1768 stars are EBs. 
 Information on these EBs is presented in Tables~\ref{tab_gen_n} and ~\ref{tab_gen_o}. Coordinates and mean $B_{EROS}$, $R_{EROS}$  magnitudes %and colours
  are taken from the EROS-2 catalogue. %, {\bf but we rounded the latter to two digits.
   Periods also generally are those in the EROS-2 catalogue, however an asterisk marks those stars which %but a note is added for stars with 
  period was recalculated in this work.%Ida's comment
  Example light curves are shown in Figure~\ref{lc_examples} with additional information about periods and number of  data-points.  
  \begin{figure}
  \includegraphics[width=8cm]{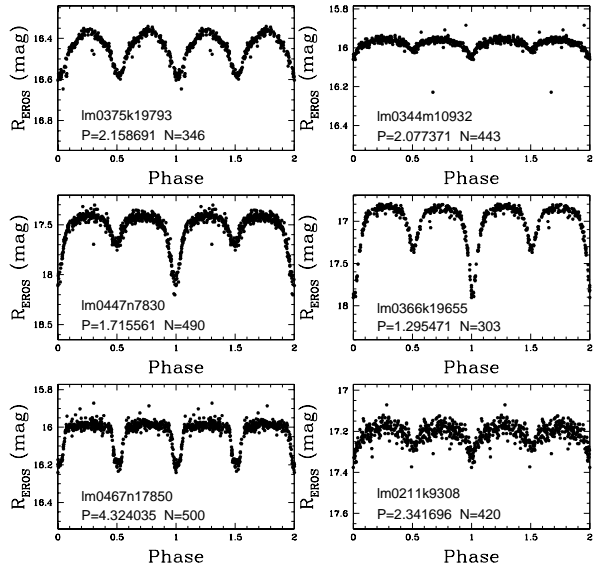}
  \caption{Examples of the $R_{EROS}$ light curves for  LMC EBs detected in the EROS-2  candidate CCs dataset. P - period (days), N - number of observations.} %not 99.99
  \label{lc_examples}
\end{figure}
%I removed the figure because of the comment of Vincenzo.
%Figure~\ref{err} shows the mean errors of the EB photometry in  the $R_{EROS}$ and $B_{EROS}$ passbands. They were derived as the average of the single-epoch errors provided in the EROS-2 catalogue. 
  $B_{EROS}$ and $R_{EROS}$ time-series photometry of the EBs is provided in Table~\ref{EROS_phot.  More details on the content of these and other tables in the paper and specifically on the accuracy/precision of the various entries is provided in the Appendix section.}

%\begin{figure}
 % \includegraphics[width=8cm]{Err_BR.ps}
%  \caption{Mean photometric errors of the EB photometry in the $R_{EROS}$ and $B_{EROS}$ passbands.\label{err}}
%\end{figure}

\begin{figure}
  \includegraphics[width=8cm]{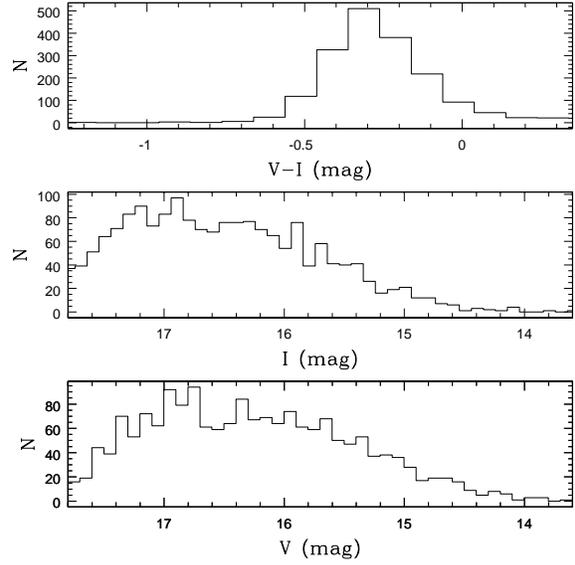}
  \caption{Distributions of mean $V, I$ magnitudes and $V-I$ colours of our sample of EBs.}
  \label{VIhist}
\end{figure}

Figure~\ref{VIhist} shows histograms of the distributions of mean $V, I$ magnitudes and $V-I$ colours for our sample of EBs.  The mean  $V$ and $I$ magnitudes  range from $\sim$17.8 to $\sim$13.2 mag  (which obviously reflect the initial cuts in magnitude used to extract  the sample of candidate CCs from the EROS-2 catalogue) and from $\sim$18.1  to $\sim$13 mag, respectively,  with a peak around 17.1-17.2 mag in both bands. The $V-I$ colours range from $\sim-1.3$ to 0.33 mag and peak at $V-I = -$0.3 mag,  which  reflects instead the colour selection we applied in the CMD of Figure~\ref{cmd_eros} to separate binaries from bona-fide CCs. 
Hence, according to their blue colours  the EBs in our sample are mainly composed of hot components (MS stars or blue giants), for this we have classified them as ``hot eclipsing binaries" (HEBs). 
%{\bf It is remarkable how effective the selection of  candidate CCs only based on the $PL$ relation was in producing also an almost pure sample of HEBs.
%  by simply selecting candidate CCs based on their $PL$ relation. In fact, extrapolating from \citet{Mor2014}  findings we expect only 

 It is remarkable the effectiveness of the candidate CCs selection based only on the PL relation in producing also an almost pure sample of HEBs. Indeed, by extrapolating the  \citet{Mor2014} findings we expect only a few (about 50) further EBs with $V-I$ colour between 0.33 and 1.677 mag to still contaminate the CCs sample. 
On the other hand, we cannot estimate the degree of completeness of our LMC HEB sample since we only had access to the portions of the EROS-2 catalogue 
corresponding to the candidate Cepheids and RR Lyrae stars. However, judging from the right panel of  Moretti et al.'s   Figure 6 the EROS-2 catalogue contains a vast number of objects with $B_{EROS} - R_{EROS} <$ 0.2 mag. We suspect  that many of them are HEBs.
%according to the o not have access to the whoHowever, we are not able to check do not have available the entire catalogue }
%The peaks  $V-I$, $V$ and $I$ magnitudes due to the selection effect.}
 
\begin{figure}
  \includegraphics[width=8cm]{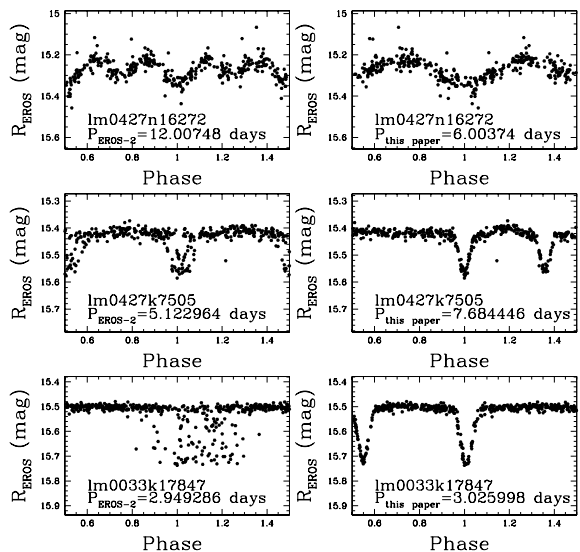}
  \caption{Light curves of EBs before (left panel) and after (right panel) correction of the period (see text for details).}
  \label{change_per}
\end{figure}
We have compared the periods provided by  EROS-2 to those determined by the visual inspection of light curves with GRATIS ($P_{GRATIS}$). For the majority of binaries $P_{GRATIS}$ is in  good agreement with $P_{EROS}$. However, in some cases, the $P_{EROS}$ was a harmonic or a subharmonic of the actual period. We corrected the period of 225 objects in the sample by multiplying $P_{EROS}$ by different constants until the shape of the light curve was consistent with that of an EB. The same technique was used by  \citet{Der2007} as part of the redetermination of periods for 3031 EBs in the MACHO catalogue.  %Richard's comment 
 Example of the light curves before and after the period correction are presented in Figure~\ref{change_per}. The systems in the middle and bottom panels of  Figure~\ref{change_per} have rather eccentric orbits which hinders the automatic determination of the period. %Carla
In some cases it was not clear if  EROS-2 determined aliases of the true period or if the binary star had only one strong expressed minimum. For these objects we decided to use the periods provided by EROS-2.

%We also compared our corrected periods with the periods from other catalogues of EBs (MACHO, OGLE~III, OGLE~IV). Among the 225 objects for which we corrected the period 163 were also detected by the OGLE~III survey, and our corrected periods are in good agreement (to within 1\%) for all but one system. % For one object  lm0030n12500 we determined  a new period which is one forth of the period provided by EROS-2, whereas OGLE~III provided a period approximately equal to 2/3 of the %EROS-2 period. 
 %Other 19 objects with corrected periods were observed  by the MACHO survey (\citealt{Facc2007}, \citealt{Der2007}).  For all of them the MACHO's periods are in agreement with our $P_{GRATIS}$. Additional 6 objects were found in the OGLE~IV catalogue. Their periods are in a good agreement (difference less than 1\%) with the periods determined with GRATIS.
 %Another 6 and 19 objects with corrected periods were observed by the OGLE~IV and MACHO (\citealt{Facc2007}, \citealt{Der2007}) surveys, respectively. The corrected periods for all of these are in good agreement with the published values (to within 1\%).
 %In conclusion, of the 225 objects  for which we corrected the periods, 188 were detected by other surveys and our estimates are confirmed in all but one of these cases (i.e>99\%). 
 %These previously determined periods support our corrected periods for 187 of them. Hence, our  corrections are confirmed in 99\% of the cases.  
 
We studied the period distribution of our sample of EBs even though the true periods of these objects cover  a relatively narrow range (from $\sim0.89$ to $\sim20$ days). 
The distribution of periods is shown in Figure~\ref{Pdist}. 
Most of our EBs are short-period systems. The distribution sharply peaks between 1 and 2 days and the majority of EBs in our sample (94\%) have periods shorter than 5 days.
 %This result is consistent with those found by other microlensing surveys. In particular, the OGLE~III period distribution based on the sample of  26121 EBs in the LMC peaks around 2.5 days (\citealt{Grac2011}).   \citet{Der2007} analysed the period distribution of 3031 EBs from the MACHO database and found that the majority of them have short periods, peaking between 1 and 2 days, and only approximately 20\% of the systems have periods longer than 10 days. Similarly, for their sample \citet{Facc2007} found a period distribution peaking strongly in the 0.8-4 day interval, with a tail in the 10-100 day range,  of EBs from the MACHO catalogue.

\begin{figure}
 \includegraphics[width=8cm]{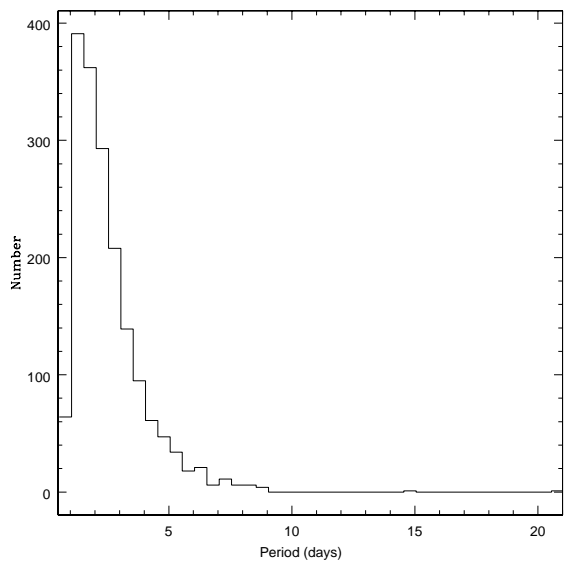}
 \caption{Period distribution of our sample of LMC EBs.}
\label{Pdist}
\end{figure}

\subsection{Cross-correlation with other catalogues of EBs in the LMC\label{cross}}
We cross-correlated our sample of 1768 HEBs with the catalogues of EBs identified  in the LMC by other microlensing surveys. Specifically, we considered:  the first stage of the EROS survey (\citealt{Gris1995}),  the MACHO survey (\citealt{Al1997, Der2007, Facc2007}), the OGLE~III (\citealt{Grac2011})  and~IV (\citealt{Sos12})  surveys. Objects in the various  catalogues were cross-identified when their right ascension and declination differed by less than $10\arcsec$,  and the periods differed by less than 1\%. We also considered objects located within less than $10\arcsec$ and with the ratio of the periods approximately equal to integer numbers, in case one of the surveys had picked harmonics or subharmonics  of the true period. We used a 
rather large pairing radius in order to avoid missing counterparts of our EBs in other catalogues, however, we note that the vast majority of the  counterparts were found to be within a pairing radius of $1\arcsec$ (OGLE~III: 99\%;  MACHO from \citealt{Facc2007}: 57\%;  MACHO from \citealt{Der2007}: 63\%;  EROS: 100\%).

Twenty-five out of seventy-nine EBs detected in the LMC bar by the first stage of  the EROS microlensing  survey (\citealt{Gris1995}) have a counterpart  in our sample of  HEBs.
%The first stage of  the EROS microlensing  survey, which was carried out from 1991 to 1992, detected 79 eclipsing binary stars located in the bar of the Large Magellanic Cloud \citep{Gris1995}. We cross-correlated EROS sample with our sample of EBs and discovered that 
  % binaries have been already detected during the first stage of the EROS survey. 
%The bottom-left
 Panel (a) of  Figure~\ref{cat} shows the position of those 25 EBs (green dots) on the map of our 1768 HEBs (black dots). %and those of them that have been detected during the first stage of the EROS survey (green dots).
\begin{figure}
  \includegraphics[width=8cm]{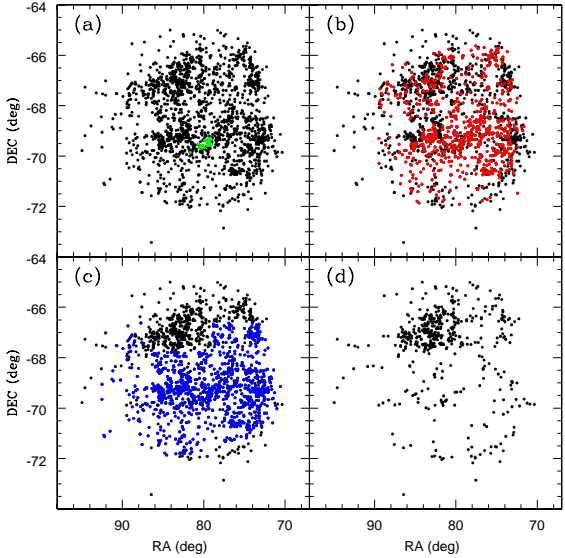}
  \caption{Panels (a)-(c): Spatial distribution of the 1768 HEBs analysed in this study (black dots) compared with those detected from previous surveys; (a) the first stage of the EROS survey (green dots), (b) the MACHO project (red dots), (c) the OGLE III/IV surveys (blue dots). Panel (d) shows the location of the 493 new EBs detected only by the EROS-2 survey.}
  %Spatial distribution of HEBs in the LMC. In all panels the black dots identify the 1768  HEBs analyzed  in the present study. Marked in different colours are the EBs already detected by other surveys, namely:  the first stage of the EROS survey (green dots panel (a));  the MACHO project (red dots in panel (b));  the OGLE~III and IV surveys (blue dots in panel (c)). Panel (d) shows the location  of the 493 new EBs detected only by the EROS-2 survey.} 
  \label{cat}
\end{figure}
%
%MACHO is a microlensing survey carried out observations between 1992 and 2000 with the 1.27 m Great Melbourne Telescope at Mount Stromlo Observatory, Australia. 
%%%The MACHO survey identified an initial sample of 611 EBs \citep{Al1997} in the LMC. Subsequently, \citet{Der2007} reanalysed the  eclipsing variables in the MACHO database, %%%corrected their periods  and presented a ``clean" sample of 3031 EBs. \citet{Facc2007} provided a new sample of 4634 EBs in the LMC from the MACHO catalogue, expanding the previous %%%sample of 611 objects from \citet{Al1997}. 
 The cross-correlation with  \citet{Der2007} and \citet{Facc2007} catalogues of EBs detected in the LMC by the MACHO survey shows that 797 objects were already known 
(panel (b) of Fig.~\ref{cat}).  
%Figure~\ref{cat} (upper-left panel) shows the position of eclipsing binaries from our sxample (black dots) and those of them that have been detected by the MACHO survey (red dots).
%Finally, OGLE~III and IV ) is the survey designed to search for microlensing events, carrying out the observations with the 1.3-m Warsaw telescope at Las Campanas Observatory in Chile 
%\citep{Grac2011}. During  the third phase OGLE~III, performed from July 2001 till  May 2009, the survey covered 40 square degrees in the LMC. Observations were taken in the standard I and  %photometric bands.    
The cross-correlation with the sample of 26121 EBs from the OGLE~III catalogue (\citealt{Grac2011}),  the 1377 EBs and the 156 ellipsoidal stars in the OGLE~IV catalogue (\citealt{Sos12}) showed that 1074 objects  were already known (panel (c) of Fig.~\ref{cat}). %, in blue in the electronic edition of the journal. % from the EROS-2 sample have counterparts in the OGLE~III catalogue. 
%Figure~\ref{cat} shows  position of eclipsing binaries from our sample (black dots) and those of them that have been detected by the OGLE~III survey (blue dots). 
We also cross-matched our sample with the spectroscopy of massive stars available from the VLT-FLAMES surveys of  Evans et al. (2006,2011), in the NGC\,2004, N11 and 30~Doradus regions of the LMC. Eight objects from our sample have been observed by these surveys: four stars in 30~Doradus, two in NGC\,2004 and two in N11, as summarized in Table~\ref{flames}. Four of these eight objects have a counterpart in the OGLE~III catalogue, one was observed by the MACHO project, three objects have not been detected before. Optical spectroscopy is available for a further five of our detected EBs,  from observations with the AAOmega multi-object spectrograph on the Anglo-Australian Telescope, one of these objects has not been detected by previous surveys.

To summarize, a total amount of 1275  sources in our EROS-2 HEBs sample had previously been detected by other surveys (OGLE~III, OGLE~IV, MACHO, EROS, with the FLAMES and AAOmega spectrographs), whereas 493 are new identifications.
%are new s from the EROS-2 sample are new identifications. 
The positions of these new discoveries  in the LMC are shown in panel (d) of Figure~\ref{cat}. As expected they are mainly located in the outer regions of the LMC.

We also compared our corrected periods with the periods from other catalogues of EBs (MACHO, OGLE~III, OGLE~IV). Among the 225 objects for which we corrected the period 163 were also detected by the OGLE~III survey, and our corrected periods are in good agreement (to within 1\%) for all but one system.
% For one object  lm0030n12500 we determined  a new period which is one forth of the period provided by EROS-2, whereas OGLE~III provided a period approximately equal to 2/3 of the %EROS-2 period. 
 %Other 19 objects with corrected periods were observed  by the MACHO survey (\citealt{Facc2007}, \citealt{Der2007}).  For all of them the MACHO's periods are in agreement with our $P_{GRATIS}$. Additional 6 objects were found in the OGLE~IV catalogue. Their periods are in a good agreement (difference less than 1\%) with the periods determined with GRATIS.
 Another 6 and 19 objects with corrected periods were observed by the OGLE~IV and MACHO (\citealt{Facc2007}, \citealt{Der2007}) surveys, respectively. The corrected periods for all of these are in good agreement with the published values (to within 1\%).
 In conclusion, of the 225 objects  for which we corrected the periods, 188 were detected by other surveys and our estimates are confirmed in all but one of these cases (i.e. $>99\%$). 
 %These previously determined periods support our corrected periods for 187 of them. Hence, our  corrections are confirmed in 99\% of the cases.  

Figure~\ref{P_ogle} shows the comparison of the periods used in this paper with those provided by the OGLE~III and OGLE~IV catalogues for the 1072 objects in common. 
%with us.
%have a counterpart in the OGLE~III catalogue. 
Two objects, namely  lm0185l23772 and lm0030n12500 are not shown in the plot because their OGLE~III periods are harmonics of the periods derived in this paper so they differ significantly. We checked the light curves of these objects with GRATIS and could not confirm the periods in the OGLE~III catalogue. In particular, for lm0185l23772 (OGLE-LMC-ECL-09445) we confirmed the period provided by the EROS-2 catalogue (P=4.97 days), whereas for lm0030n12500 (OGLE-LMC-ECL-20762) we determined a new period (P=1.4998 days) which is one third of the period provided by EROS-2, while OGLE~III determined a period approximately equal to two thirds of the EROS-2 period.  The light curves of these objects are presented in Figure~\ref{lc}. %JB comment
Apart from these two objects,  the periods adopted in this paper and those in the OGLE~III generally differ by less than 0.03\% (see Figure~\ref{P_ogle}).

\begin{figure}
  \includegraphics[width=8cm]{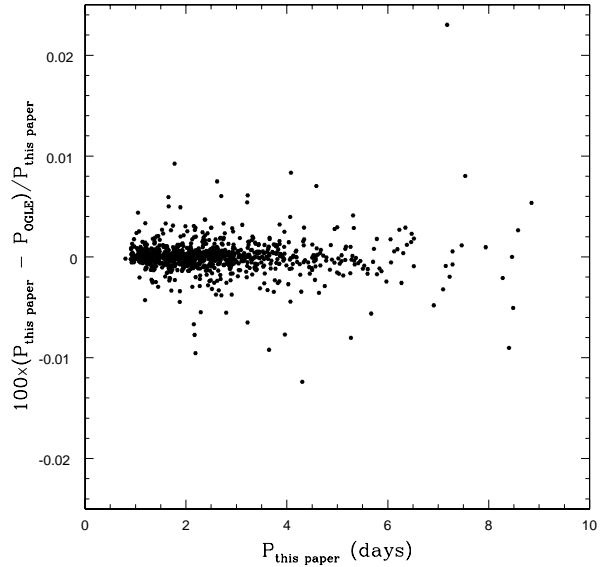}
  \caption{Comparison between periods adopted in this paper and those in the OGLE~III and OGLE~IV catalogues for the 1072 EBs in common. Two objects, namely lm0185l23772 and lm0030n12500, were not included because their  OGLE periods differ significantly from our values (see text and Figure~\ref{lc} for details).} 
  \label{P_ogle}
\end{figure}

\begin{figure}
  \includegraphics[width=8cm]{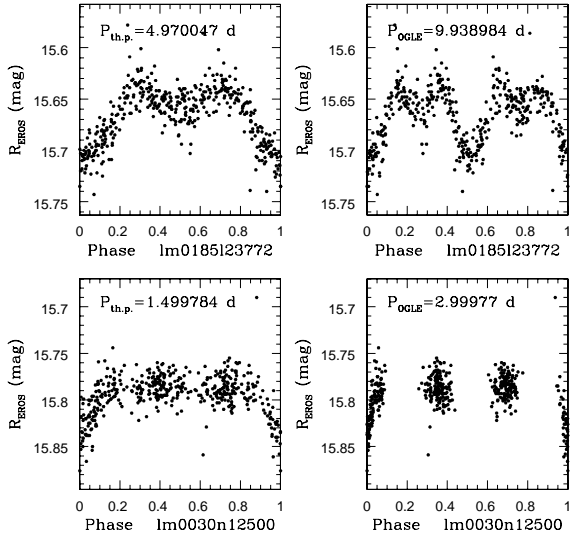}
  \caption{Light curves of lm0185l23772 and lm0030n12500 with the periods used in this paper (left panel) and provided by the OGLE~III catalogue (right panel).} 
  \label{lc}
\end{figure}

\subsection{Characteristics of EBs with existing spectroscopy}
\subsubsection{Cross-matches with the VLT-FLAMES surveys}

As already mentioned in Section~\ref{cross}, eight HEB systems in our sample have  existing optical spectroscopy from surveys  with FLAMES  at the VLT (\citealt{Ev2006}, \citealt{Ev2011}), as summarized in Table~\ref{flames}.  All were detected as binaries in the multi-epoch spectroscopy, except for VFTS\,462 (Dunstall et al. in prep).  In addition to the EROS-2 periods, estimates are also available from the OGLE~III data for the four VFTS systems (Graczyk et al. 2011), with excellent agreement in all cases; the other four systems (in  NGC\,2004 and the N11 region) are beyond the OGLE~III survey area. 

Quantitative analysis of the VFTS spectra is still underway, but evolutionary mass estimates (of the primaries) of the other systems are available from Hunter et al. (2008); $M$\,$=$\,13\,$M_{\odot}$ for both N11-107 and N11-119, and $M$\,$=$\,11 and 10\,$M_{\odot}$ for NGC\,2004-079 and NGC\,2004-094, respectively\footnote{However, note that these estimates were on the basis of effective temperatures  adopted from the spectral classifications, and the expected uncertainties on these masses is typically 30\% \citep{Hunt2008}}.  Photospheric chemical abundances were presented for the two systems in NGC\,2004 by \citet{Hunt2009}, with seemingly unremarkable nitrogen abundances.
The spectroscopy from the FLAMES surveys was effective in detecting spectroscopic binaries, but further monitoring is generally required to characterize the orbital parameters (e.g., \citealt{Rit2012}).  Indeed, spectroscopic monitoring of a subset of the O-type binaries discovered by the VFTS is now underway (P.I. Sana), and includes VFTS 061 among its targets.

\subsubsection{AAOmega spectroscopy}
Optical spectroscopy is available for a further five of our detected EBs, from observations with the AAOmega multi-object spectrograph on the Anglo-Australian Telescope, obtained during 2006 February 22-24 (P.I. van Loon). The five targets discussed here were obtained as part of two fields 
centred on N11 and 30~Dor.  AAOmega is a twin-arm spectrograph (providing simultaneous blue/red coverage), but only the blue data are discussed 
here.  Both fields were observed on the first night with the 1700B grating and two central wavelengths (4100 and 4700\,\AA), giving coverage of 3765-5015\,\AA, at a resolution of 1\,\AA.  The 30~Dor field was also observed on the second night with the 1500V grating, at a central wavelength of 4375\,\AA, providing coverage of 3975-4755\,\AA, at a resolution of 1.25\,\AA.  These data were reduced using the AAOmega 
reduction pipeline and the relevant spectra were rectified and co-added.

Spectral classifications for the five systems are presented in Table~\ref{flames}, in which we have employed the same framework as that used by Evans et al. 2014 (in prep.).  All five systems have early B-type spectra (in line with the expectation of these as HEBs), with morphological evidence for binarity (double-lined and/or asymmetric profiles) in all but one.

%\tablefoot{E06 \citep{Ev2006}; W14 \citep{Wal2014}; E14 (Evans et al. in prep). OGLE~III periods from \citet{Grac2011}}

\section{Classification of eclipsing binaries}\label{fourier}

 Our classification of the EBs was based on both the Fourier analysis (\citealt{Ruc1993},1997 and \citealt{Mac1999}) and the visual inspection of the light curves. 
 \citet{Ruc1993} showed that a simple description of the light variation of contact binaries could be obtained through the cosine Fourier decomposition $\sum a_{i}\cos(2\pi i\phi)$. \citet{Ruc1997} performed the Fourier analysis of the $I_C$-band  light curves to extract a sample of contact binaries from OGLE  EBs in nine fields in Baade's Window. 
 In this study we use an updated version of this method, developed by one of us (C.M). We fitted the $R_{EROS}$ light curves of  the 1768 HEBs  in our sample with a model light curve consisting of six Fourier terms: 
\begin{equation}
\sum_{i=0}^4 a_{i}\cos(2\pi i\phi) + a_{5}\sin(2\pi \phi)
\end {equation}
 
In equation 3 the coefficient $a_{0}$ is the average magnitude of the model fit,  $a_{1}$ and $a_{3}$ represent the difference in depth between two eclipses, the sine term $a_{5}$ is related to lack of symmetry between maxima, 
 %It goes to zero when the minima have equal depths. 
 $a_2$ reflects the total amplitude of the binary variability and $a_4$ is related to  the eclipse ``peakedness" that goes to zero for the light curves of contact binaries. Hence, the combination of the two cosine coefficients $a_2$ and $a_4$ could serve as a separator of contact and non-contact binaries. Namely,  the curve described by the relation $a_{4}=a_{2}(0.125-a_{2})$, where both coefficients are negative,  separates the  regions of the contact and non-contact binaries on the $a_{2}$ versus $a_{4}$ plane \citep{Ruc1993}.   However,  as mentioned in the Introduction, not only genuine contact binaries, but also variables with contact-like light curves such as the ellipsoidal variables might be misclassified as contact binaries. Ellipsoidal variability is observed in close binary systems when one or both components is (are) distorted by the tidal interaction with the companion. The main reason of the variability is the change of the projected areas on the sky because of the orbital motion of the components. Large samples of ellipsoidal variables were published by \citet{Sosz2004}, who used OGLE~II and OGLE~III photometry, and by \citet{Der2006}, who used the MACHO database. Since the light curves of contact binaries and  non-eclipsing  ellipsoidal variables have similar shapes and could be easily mistaken, we adopt the term ``contact-binary-like" systems for all objects passed by the Fourier filter.

By analysis of the Fourier decomposition of the light curves in the $R_{EROS}$  passband  we identified the contact-like binaries in our sample. The light curves were not expressed in magnitudes but in intensity units,  relative to the maxima at phases in the range [0.24, 0.26]. %Richard, ask Carla
Columns from 8 to 13 of Tables~\ref{tab_gen_n} and~\ref{tab_gen_o}  present the 6 Fourier coefficients $a_0$ to $a_5$ of the Fourier analysis for the 493 new HEBs and for the 1275 HEBs already detected by other surveys, separately.
% We fitted the $R_{EROS}$ light curves of  the 1768 HEBs  in our sample with a model light curve consisting of six Fourier terms: {\bf $a_{0}$ (related to the average magnitude),  $a_{1}$-$a_{4}$ (cosine terms)  and  $a_{5}$ (sine term). Only cosine terms $a_2$ and $a_{4}$ are being used to build the Fourier light-curve shape filter \citep{Ruc1993}.} 
%: 6 Fourier coefficients $a_0-a_5$. 
%However,  as mentioned in the Introduction, not only genuine contact binaries, but also ellipsoidal variables with contact-like light curves might be misclassified as contact binaries. Ellipsoidal variability is observed in close binary systems when one or both components is (are) distorted by the tidal interaction with the companion. The main reason of the variability is the change of the projected areas on the sky because of the orbital motion of the components. Large samples of ellipsoidal variables were published by \citet{Sosz2004}, who used OGLE~II and OGLE~III photometry, and by \citet{Der2006}, who used the MACHO database. Since the light curves of contact binaries and  non-eclipsing  ellipsoidal variables have similar shapes and could be easily mistaken, we adopt the term ``contact-binary-like" systems for all objects passed by the Fourier filter.
 Figure~\ref{fit} depicts examples of the resultant fits for both contact-like (lower panel) and non-contact (upper panel) binary systems. 
It should be noticed that six harmonics generally allow very satisfactory fits for contact-like systems, whereas some noticeable differences arose between the observations  and the fitted curves for non-contact binaries, which would indeed require a much larger number of harmonics (8-10 or more) to be modelled. %Andres
This is often due to elliptical orbits, yielding a shift of the secondary minimum from phase 0.5 of the non-contact systems.
%(e.g, the system shown in the upper panel of Figure~\ref{Fit_ex}), 
%which  would require a larger number of harmonics to be reproduced.

 Figure~\ref{a2_a4} shows the position of 1768 EBs from our sample in the $a_{2}$ versus $a_{4}$ plane. The solid line in the figure shows the contact locus line defined by 
 \citet{Ruc1993}.  We classified objects located below the line as contact-like binaries (324 sources) and those above as non-contact binaries (1444 sources). 
 However, being aware that detached and semi-detached systems could accidentally appear below the locus line due to a bad fit of the light curve, we  visually inspected 
the light curves of all the objects (324 stars) located below the line; we discovered  eight objects which have light curves without the characteristic form of contact binaries, so we discarded them.
% We visually 
\begin{figure}
  \includegraphics[width=8cm]{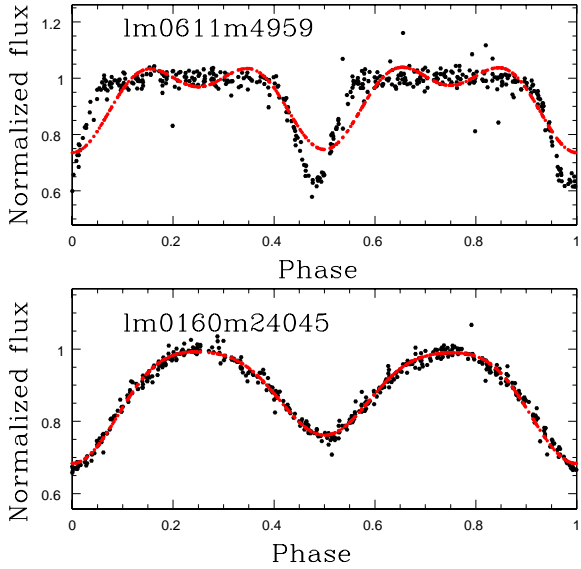}
  \caption{Examples of the Fourier fit obtained using 6 harmonics to model the light curve of detached (upper panel) and contact-like (bottom panel) binaries in our sample. Black dots represent the observational data, red solid lines show the resultant %(in red in the electronic edition of the journal) 
  Fourier fits. Six harmonics are clearly not sufficient to reproduce detached systems.} %Andres
  \label{fit}
\end{figure}
\begin{figure}
  \includegraphics[width=8cm]{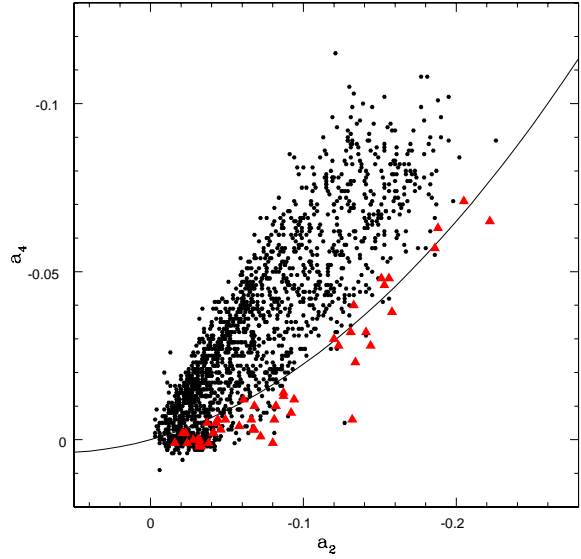}
  \caption{Fourier coefficients $a_2$ and $a_4$ of the 1768 HEBs in our sample. The solid curve is described by the relation $a_{4}=a_{2}(0.125-a_{2})$ which, according to  \citet{Ruc1993} separates the regions of contact and non-contact binaries. Objects located below the line are considered to be contact systems.  Red filled triangles %(in red in the electronic edition)
   identify objects classified as contact binaries in the OGLE~III catalogue (see text for details).} 
  \label{a2_a4}
\end{figure}
% inspected the light curves of the 324 stars classified as contact-like binaries  %As a result, our sample of ECs reduced to 315. 
 In conclusion,  in our analysis a system is classified as contact-like if it is located  below the locus line from \citet{Ruc1993}  on the $a_{2}$ versus $a_{4}$ plane and its light curve has the characteristic shape of a contact system.

We compared our classifications with those from the OGLE~III catalogue. Out of 1055 objects in common, 48 stars were classified as contact systems in the OGLE~III catalogue (red filled triangles in Figure~\ref{a2_a4}).  Figure~\ref{a2_a4} shows that the  majority of these objects 
%that were classified as ECs by the OGLE~III team 
 are in fact located below the locus line traced by \citet{Ruc1993}, while the majority of the systems classified as non-contact variables by OGLE~III   are above the curve. However,  eight objects  located marginally  above  the locus line  were  classified as contact binaries by OGLE~III. We checked their light curves and confirmed that these binaries are indeed contact-like systems.  When including these, the final number of contact-like binaries in our sample is 324.
%On the contrary, 50 out of 1055 objects in common between OGLE~III and our sample of HEBs were classified as contact-like binaries by us and as detached, semi-detached or ellipsoidal systems by OGLE III. 
In contrast, 50 of 1055 objects in common were classified as contact-like systems by us, but as detached, semi-detched or ellipsoidal systems by OGLE~III.
We double checked their light curves, and found that  our classification is in disagreement with OGLE~III in some cases. The majority of these objects have low amplitudes so it is difficult to provide an exact classification by visual inspection of light curves. In the following analysis we use our classification for those objects, thus our final sample consists of 324 contact-like binaries and 1444 non-contact systems.

\section{Period-Luminosity relation of eclipsing binaries}
\subsection{PL-relation of EBs from the EROS-2 sample}\label{er_pl}
The $PL$ relation of blue, luminous contact systems, observed in the LMC by the MACHO project, was studied by \cite{Ruc1999}. He suggested the existence of a $PL$ relation 
at  maximum light in the visual band, but with a large scatter, possibly due to unaccounted effects of the interstellar extinction. 
Following  \cite{Ruc1999} we have investigated whether our sample of HEBs follows  a $PL$ relation at  maximum light using the red passband photometry of EROS-2 ($R_{EROS}$) and 
near-infrared photometry in the $K_{\rm s}$-band obtained as part of the VMC survey (\citealt{Cioni2011}).  
The latter was used in order to minimize possible extinction effects.%Andres
%We expect that any problems due to extinction should be minimized using the red passbands. 
\begin{figure*}
  \includegraphics[width=16cm]{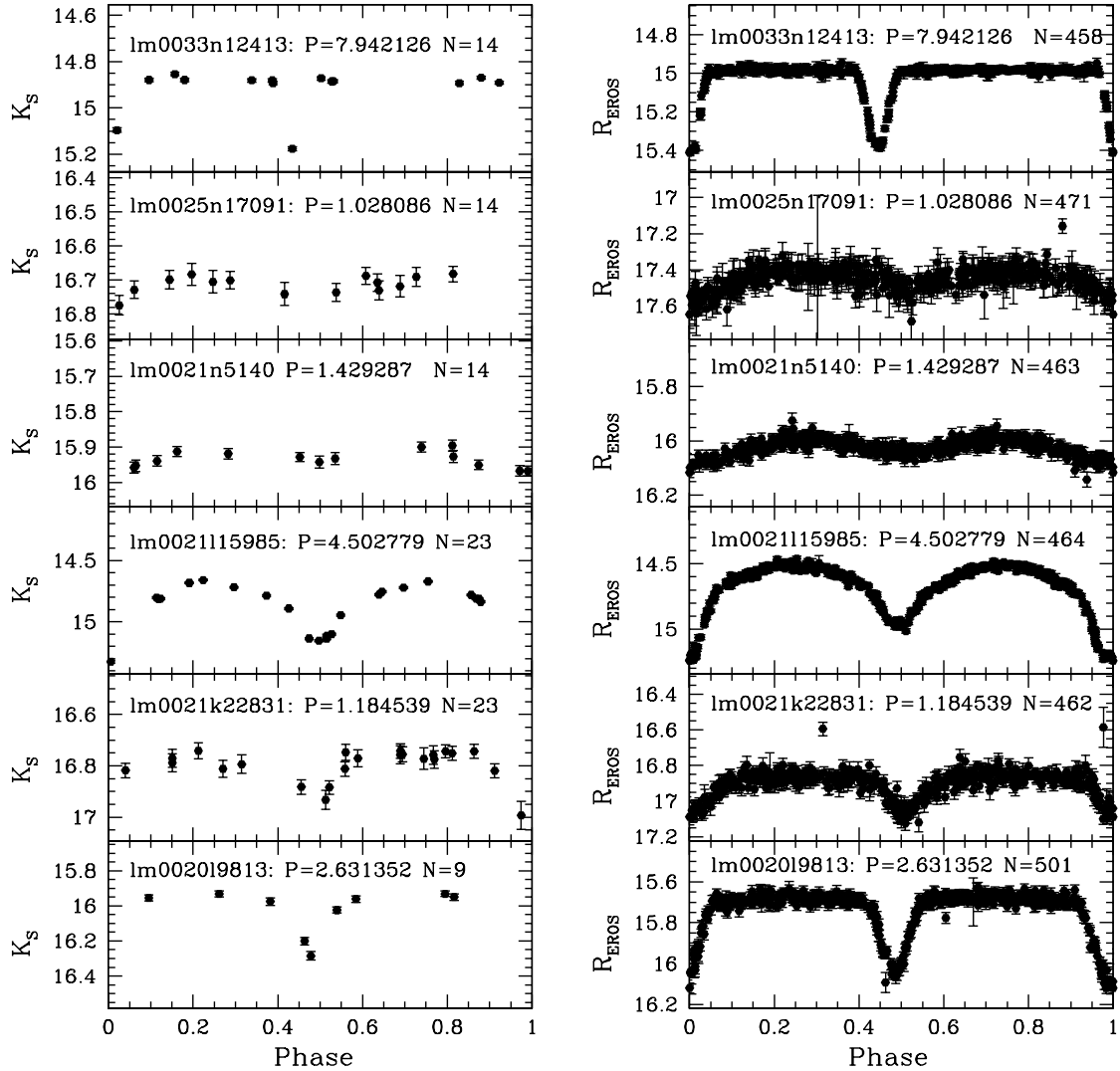}
 % \qquad
   \caption{Light curves in the  $K_{\rm s}$ (left panels) and $R_{EROS}$ (right panels) passbands of example HEBs with a counterpart in the VMC catalogue. P - period (days), N - number of observations in the corresponding passband.} %JB comment
  \label{Klc}
\end{figure*}
Started in 2009, the VMC survey aims at covering a total area of 116 $\deg^2$ in the LMC with 68 contiguous tiles. The $K_{\rm s}$-band observations are taken in time-series mode over 12 (or more) separate epochs. Each single epoch reaches a limiting  $K_{\rm s}$ band magnitude $\sim19.3$ mag (see Figure 1 of \citealt{Mor2014}). On the bright side, VMC is limited by saturation which produces an essential departure from linearity starting at $K_{\rm s}$ $\sim10.5$ mag. 
 Aperture photometry of the VMC images is performed by the Cambridge Astronomical Survey Unit (CASU) through the VISTA Data Flow System (VDFS) pipeline.
The reduced data are then sent to the Wide Field Astronomy Unit (WFAU) in Edinburgh where they are catalogued by the Vista Science Archive (VSA; \citealt{Lewis2010,Cross2012}).
%images. %and computes the Julian Day (JD) of observation for each source by averaging only the JDs of the pawprints in which that source was observed. 
%The single epoch reduced data are then sent 
To date, the complete multi-epoch dataset is available for ten  LMC tiles, whereas a further seven tiles have been observed at least once (observations completed until the 1st of April 2013).  
These 17 tiles sample different regions of the LMC from the inner bar to the outer regions. 

We have cross-matched our catalogue of 1768 HEBs against the VMC catalogue  available at VSA (observations completed until the 1st of April 2013) %Ida's comment
 for the 17 tiles and found  999 binaries in common using a pairing radius of 1$^{\prime \prime}$.   %Andres
%We analysed $K_{\rm s}$-band light curves of binary stars from the sample observed by the VISTA for Magellanic Clouds survey. This is ESO public survey obtaining YJ$K_{\rm s}$ photometry ($
%\lambda=1.02$, 1.25 and 2.15 $\mu$m, respectively) of the Magellanic Clouds System  \citep{Cioni2011}. The observations are performed with the 4-m near-infrared optimized VISTA %telescope. The VMC covers 116 $deg^2$ in the LMC, 46 $deg^2$ in the SMC, 20 $deg^2$ in the Bridge and 3 $deg^2$ in the Stream. 
%There are 999 binaries in our sample which have counterparts in the VMC catalogue. 
Examples of the  $K_{\rm s}$ and $R_{EROS}$  light curves for some of these binaries are shown in Figure~\ref{Klc}. The  $K_{\rm s}$ time-series photometry is provided in Table~\ref{VMC_phot}. 
The number of phase-points of the $K_{\rm s}$-band light curves varied from a minimum of one for EBs  located in tiles with incomplete observations to a maximum of over 30 phase points 
for EBs located in regions where different tiles overlap. Furthermore,  the EBs in our sample are relatively bright sources. Therefore the shallow VMC epochs, for which the integration time of observation is half that for deep epochs, or epochs not meeting the original quality criteria (e.g.,  seeing, etc.) were enough to measure the  EBs thus increasing the number of available phase-points.

The  $R_{EROS}$ and $K_{\rm s}$ magnitudes at maximum light of the binaries that have a VMC counterpart are presented in  Table~\ref{tab_K}.  
% Figure~\ref{PReros } 
%We used these data in order to study the Period-Luminosity relation of contact eclipsing binaries in the $K_{\rm s}$ passband. 
In order %decrease the errors of 
  to better determine the $K_{\rm s}$-band magnitudes at maximum light, we performed an additional analysis of the light curves with GRATIS, for those HEBs which have 13 or more good-quality observations.  
Figure~\ref{PLReros} shows the $PL$ distribution in the $R_{EROS}$ band of the 999 EBs with a VMC counterpart, whereas Figure~\ref{PK} shows their $PL$ distribution in the  $K_{\rm s}$  band.  %of all binaries from our sample which have been observed by the VMC.
 In both figures red open circles identify the sources which we classified as contact-like systems. 
 The contact binaries for which we have 13 or more $K_{\rm s}$-band epochs (and for which maximum magnitudes were determined with GRATIS) are highlighted in green. 
 %In Fig.~\ref{PK}  we highlighted  in green the contact binaries for which we have 13 or more epochs in the $K_{\rm s}$-band light curve and the $K_{\rm s}$-band maximum magnitude was determined with GRATIS. 
 Unfortunately, the use of a  more robust method to determine the $K_{\rm s}$ magnitude at maximum light does not decrease the scatter. Both the optical and near-infrared $PL$ distributions exhibit  a very large dispersion, which is of the same order of the scatter observed in the $PL$  relation originally used by the EROS-2 team to extract the candidate CCs from the EROS-2 general catalogue of  LMC variables (see upper-right panel of Fig. 12 in \citealt{Mor2014}). Furthermore, the comparison with Fig. 12 of \citet{Mor2014} shows that the EBs are in fact responsible for the large scatter observed in the candidate CCs $PL$ in the EROS-2 bands. Thus, there does not appear to be a {\it PL} relation for HEBs.
 
\subsection{$PL$-relation of EBs from the OGLE~III catalogue} 
In order to study the $PL$ relation of contact binaries in  a more general sample and over a larger range of periods we have used the OGLE~III catalogue of EB stars published by \cite{Grac2011}.  We extracted all the objects which were classified as contact binaries  in the \cite{Grac2011} catalogue.
%We extracted from  the \cite{Grac2011} catalogue all objects which were classified as contact binaries.
 Among them we selected objects with VMC counterparts and, with both $V$ and $I$ magnitudes from OGLE~III, giving 563 objects in total. Twenty-five of these have their counterparts in our sample of HEBs from the EROS-2 catalogue and
 were already discussed in Section~\ref{er_pl}.   To account for extinction we used the LMC reddening maps derived by \citet{Has2011} on the basis of OGLE~III data. To compute extinction values in the various bands we used the relations from \citet{Sch1998} and \citet{Card1989}, these were then applied to correct each source.
 %used to study the Period-Luminosity relation of contact-like binaries in the $K_{\rm s}$ passband (Subsection~\ref{er_pl}). 

 The  reddening corrected $V_0, (V-I)_0$ CMD  of contact binaries from the OGLE~III catalogue is shown in Figure~\ref{cmd_ogle}. Contact binaries are located in two regions in the CMD: HEBs  which  contain MS stars or blue giants have    $(V-I)_0<0.3$ mag, whereas EBs with a red giant component have  $(V-I)_0\geq 0.3$  mag. The corresponding  $PL$ distributions in the  $I_0$ and  $K_{\rm s,0}$ bands are  presented in Figures~\ref{PI_ogle} and ~\ref{PK_ogle}, respectively. In the figures HEBs are indicated with black dots and binary systems containing red giants are indicated with  red triangles. 
For 164 EBs with a red giant component,  which have 13 or  more good-quality epochs from the VMC survey, we analysed the light curves with GRATIS in order to determine the $K_{\rm s}$ magnitude at maximum light with a good accuracy. In Fig.~\ref{PK_ogle} we have highlighted these objects  with green triangles. While contact HEBs from the OGLE~III sample do not distribute along a $PL$ sequence, contact binaries containing red giant components seem to follow at least one, maybe two,  different $PL$ sequences. 

To further investigate this point we restricted our analysis to 164 objects with carefully determined $K_{\rm s}$ maximum magnitudes (green triangles in Fig.~\ref{PK_ogle}).  Their {\it PL} distribution is shown in Figure~\ref{PLs_cont}. As it could be seen, there are 11 objects with low periods which do not follow any {\it PL} sequence (blue open circles in Fig.~\ref{PLs_cont}). We checked the position of these objects on the CMD (Fig.~\ref{cmd_ogle}) and found that they are located in the border region between HEBs and binaries with red giant components. Since these objects do not follow the {\it PL} sequence and could be HEBs, we discarded them from the following analysis.

For other 153 EBs  we computed  a weighted linear regression through the data by progressive discarding objects which deviate more than $3\sigma$ from the linear regression. 
%
% we showed objects with carefully determined $K_{\rm s}$ maximum magnitudes that according to the visual inspection of Figure~\ref{PK_ogle} follow Period-Luminosity relations. We determine %the coefficients of PL sequences by considering objects which are less than $3\sigma$ far from the linear regression.  
 The majority of contact systems with carefully determined $K_{\rm s}$ maximum magnitudes  (red dots in Figure~\ref{PLs_cont}) appear to follow the relation  : 
\begin{equation}\label{PL1}
K_{\rm s,0}=(-2.888\pm0.096)log(P) + (20.139\pm0.171)
\end{equation}
with rms=0.406 mag.

In Figure~\ref{PLs_cont} we have highlighted objects located more than $3\sigma$ from the {\it PL} distribution with black dots. Some of them seem to follow a {\it PL} sequence parallel to the one described by  equation~\ref{PL1} and located $\sim1$ mag fainter than the previous one.
%We checked with GRATIS  the light curves of 15 EBs which follow the PL-relation described by the equation~\ref{PL2} and didn't find any peculiarity of these objects. 

To summarize, hot contact binary systems do not follow any $PL$ relation while the existence of  red giant $PL$ sequence(s) (at least, one) seems quite clear and, as shown by  Figs.~\ref{PI_ogle} and ~\ref{PK_ogle}, this relation appears to be narrower in the $K_{\rm s}$ passband.  %Maria-Rosa'comment
However, the large scatter 
%(of about half a magnitude) even at $K_{\rm s}$  reinforces the idea that several different sequences may exist,  and along with the lack of clear, unquestionable ways to assign objects to one sequence or the other,  
make it impossible to use these sequences any further.
On the other hand, that red giants follow multiple {\it PL} relations was already reported in many studies (\citealt{Wood1999},  \citealt{Sosz2004}, \citealt{Der2006}). \citet{Wood1999} were the first to recognize five different $PL$-sequences: A, B and C, occupied by pulsating red giants, D composed by stars that have long secondary periods (LSPs),  and sequence E, containing red giants in contact eclipsing binaries and ellipsoidal variables. \citet{Sosz2004} showed that  a $PL$ relation of ellipsoidal variables could be well described by a simple model using the Roche-lobe geometry and that sequences E and D merge at specific luminosities. \citet{Der2006} presented a period-luminosity-amplitude analysis of 5899 red giant and binary stars in the LMC from the MACHO database and discovered that the $PL$ sequence of binaries is composed only by contact EBs, while detached and semi-detached systems are spread everywhere in the $PL$ plane. Moreover, they concluded that sequence E, %which comprises contact binaries and ellipsoidal variables
  is located at periods a factor two greater and overlaps with the sequence of LSPs (sequence D). In our study we confirm the existence of a $PL$ sequence containing contact binaries with red  giant components (equation~\ref{PL1}) and find evidence for a possible additional $PL$ sequence of contact binaries located $\sim1$ mag fainter.
%than the previous one (equation~\ref{PL2}).

\begin{figure}
  \includegraphics[width=8cm]{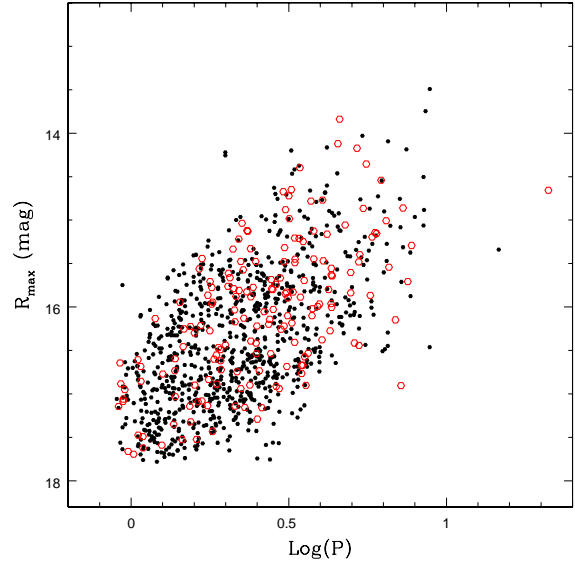}
 % \qquad
   \caption{$PL$ distribution in the $R_{EROS}$ passband of 999 HEBs  that have a counterpart in the VMC catalogue. Red open circles are objects we classified as contact binaries.} %black dots: non-contact binaries} 
  \label{PLReros}
\end{figure}
\begin{figure}
  \includegraphics[width=8cm]{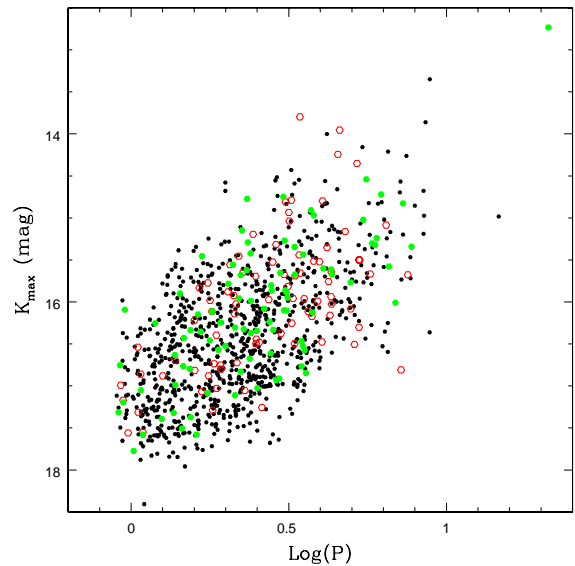}
 % \qquad
   \caption{$PL$  distribution in the $K_{\rm s}$ passband of 999 HEBs from our sample that have counterparts in the VMC catalogue. Red open circles are objects classified as contact binaries, green filled circles are 90 contact EBs for which we have 13 or more epochs in the $K_{\rm s}$ light curves.
   } %, black dots: non-contact binaries} 
  \label{PK}
\end{figure}

Furthermore, thanks to the depth achieved by the VMC data, we are able to extend the {\it PL} relation of contact binaries, containing red giants, to $K_{\rm s} \sim$ 18 mag, roughly two magnitudes fainter than in  \citet{Der2006}. 
The existence of {\it PL} relation(s) for red giant EBs and its absence for HEBs could be explained by 
intrinsic differences occurring between the two samples. 
%Firstly, our binaries with MS  components were selected in a limited range of periods, hence, the sizes of the stars do not vary significantly, while red giant binaries have periods in the range from $\sim0.3$ to $\sim300$ days.
 In the case of contact systems with red giants the total luminosity of the binary system is dominated by one component - the red giant star, the luminosity of the second component being  negligible. On the contrary, for HEBs the ratio of luminosities of the two O-B components could vary significantly. %Actually, why?
Therefore,  the scatter of the $PL$ relation of binaries with O-B components is expected to be much larger than the scatter of the $PL_{K_{\rm s}}$ relation of contact systems  with a red giant component.

\section{Summary}
\begin{figure}
  \includegraphics[width=8cm]{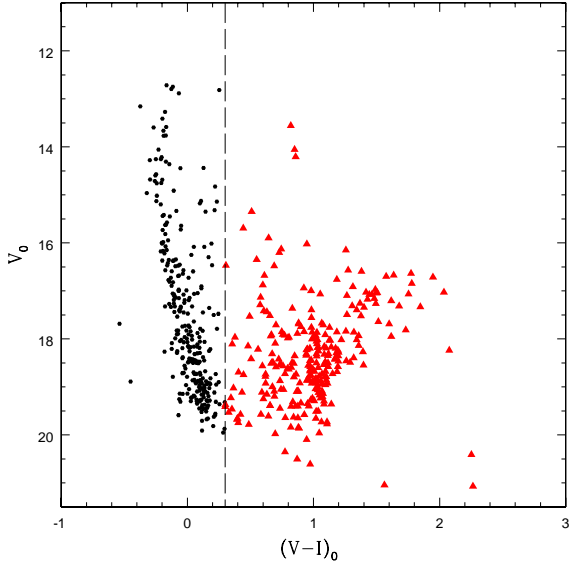}
   \caption{ Colour-magnitude diagram of  563 contact binary stars which have $V$ and $I$ (OGLE~III) and $K_{\rm s}$ (VMC) magnitudes. Black dots are objects with $(V-I)_0 <0.3$ mag and red triangles are objects with $(V-I)_0 \ge 0.3$ mag. The dashed line corresponds to $(V-I)_0=0.3$ mag.} 
  \label{cmd_ogle}
\end{figure}
\begin{figure}
  \includegraphics[width=8cm]{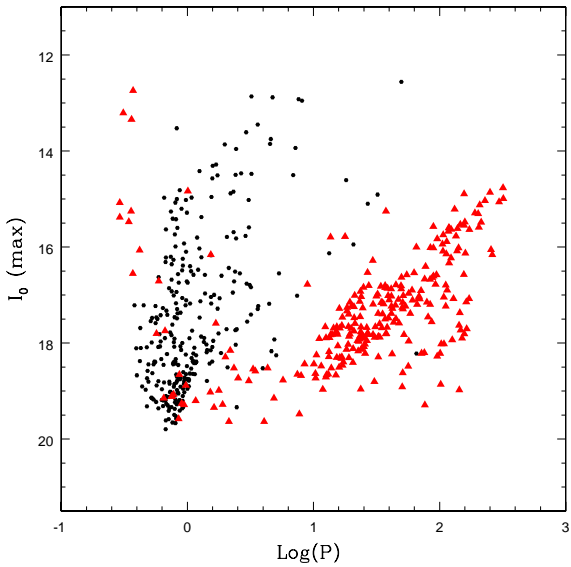}
 \caption{$PL$ distribution at $I_0$ maximum  of  the 563 contact binaries  shown in Fig.~\ref{cmd_ogle}.
%
 %from the OGLE~III catalogue which have $V$ and $I$ magnitudes provided by the OGLE~III survey and $K_{\rm s}$ magnitudes 
 %at    
%\caption{$PL$ relation in the $I$-band of  563 contact binary stars from the OGLE~III catalogue which have  $I$ magnitudes at maximum light from the OGLE~III survey and $K_{\rm s}$ magnitudes %at maximum light 
%from the VMC survey. 
Black dots are objects with $(V-I)_0 <0.3$ mag and red triangles are objects with $(V-I)_0 \ge 0.3$ mag.} 
  \label{PI_ogle}
\end{figure}

\begin{figure}
  \includegraphics[width=8cm]{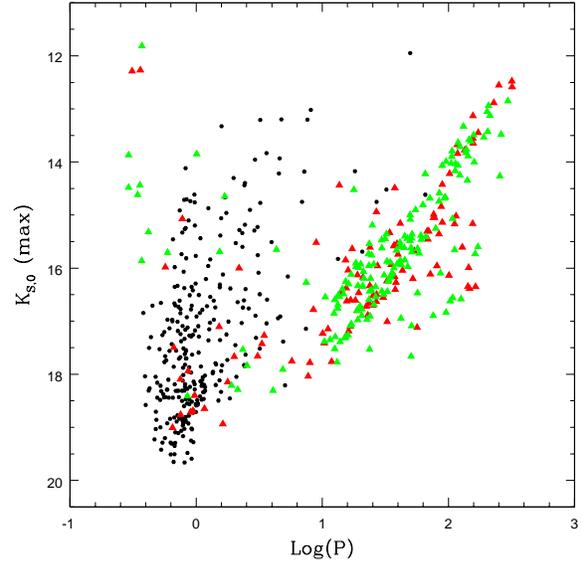}
   \caption{$PL$ distribution at $K_{\rm s,0}$ maximum of  the 563 contact binaries shown in Fig.~\ref{cmd_ogle}. 
%  
%    stars from the OGLE~III catalogue which have $V$ and $I$ magnitudes provided by the OGLE~III survey and $K_{\rm s}$ magnitudes at maximum light from the VMC survey.
     Black dots are objects with $(V-I)) <0.3$ mag, red triangles are objects with $(V-I)_0 \ge 0.3$ mag of which those with 13 or more epochs in the $K_{\rm s}$-band are marked in green.} 
  \label{PK_ogle}
\end{figure}

\begin{figure}
  \includegraphics[width=8cm]{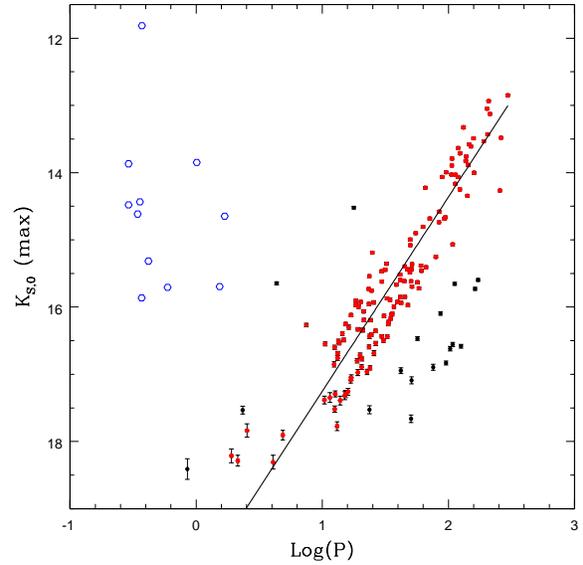}
 % \qquad
   \caption{$PL$ distribution of contact binaries containing red giant components. Blue open circles are candidate HEBs  falling in the region of binaries with red giant components (their errors are smaller than the size of the circles), red and black dots represent EBs which deviate less (red) and more (black) than $3\sigma$ from a linear regression, respectively. 
  The line is the weighted linear fit obtained from the objects marked in red.  \label{PLs_cont}}
\end{figure}

We have presented results from the analysis of 1768 EB  stars detected by the EROS-2 survey in the LMC; 493 are new identifications of eclipsing systems, generally located in the outer regions of the galaxy.  The EBs  in our sample contain hot MS stars and  blue giants, as illustrated by the spectral classifications for 13 of these binaries. %, hence we named them HEBs.
%We presented  results of the spectroscopic analysis for 13 of these binaries.
 We analysed the light curves of the 1768 HEBs and re-determined the previously-defined periods for 225 of them.  Following \cite{Ruc1993}, we divided the sample into contact-like (324) and non-contact  (1444) systems by visual inspection of their light curves and by analysis of the Fourier decomposition parameters. %, we classified 324 objects from the sample as contact systems. 
We presented $K_{\rm s}$-band light curves for  999 of these EBs  which have counterparts in the catalogue of the VMC survey. We analyzed the $PL$ relation in the optical ($R_{EROS}$ and $I$) and  $K_{\rm s}$  bands of the contact-like binaries in our EROS-2 sample and in  the more extended sample of OGLE~III  contact EBs.  Hot contact binaries do not follow $PL$ relations in the $R_{EROS}$ and $K_{\rm s}$ passbands, while binaries containing a red giant component do follow $PL$ sequences in the $I$ and $K_{\rm s}$ passbands. We computed the weighted linear regression of the {\it PL} relation in the $K_{\rm s}$ passband. There is a possible additional $PL$ sequence %of contact binaries 
 located $\sim1$ mag fainter, but the number of objects following it, is too small to allow a reliable fit. The existence of a {\it PL} sequence of contact and ellipsoidal variables down to $K_{\rm s} \sim$ 16 mag was claimed in previous studies \citep{Der2006}, in this work we extended the {\it PL} down to $K_{\rm s} \sim$ 18 mag.
 %the so-called sequence E of red giants.
% There could be several reasons to explain the  absence of a $PL$ relation for MS contact binaries. First of all, the objects in our sample cover a rather narrow range in period (from $\sim1$ to $\sim20$ day), hence the difference in size of the two MS components is not as large as for the red giants.
The luminosity ratios of the components of the HEBs (MS stars and blue giants) can vary significantly. In contrast, in contact systems with a  red giant component, the giant dominates the luminosity and the contribution from the secondary is usually negligible. As a consequence the $PL$ relation of contact HEBs can be much more scattered than the $PL$ of contact EBs with a red giant component.

 %The ratio of luminosities of the MS stars and blue giant components could vary significantly, while in contact EBs containing a red giant component the total luminosity is dominated by the red giant and the luminosity of the secondary component  is usually negligible. As a consequence the $PL$ relation of contact HEBs can be much more scattered than the $PL$ of contact EBs with a red giant component.

%We have compared the spatial distribution of HEBs and CCs in the LMC. The former  appear to be mainly located in regions of most recent star formation activity, such as 30 Doradus,  Constellation III and  the supergiant shells, hence they likely are younger objects than the CCs.
%Detected by the EROS-2 objects extend the sample of the known EBs in the Large Magellanic Cloud. 
%The location in star forming regions, the high brightness and the possibility to measure their distance with a direct, geometric technique can provide the opportunity to use the HEBs to measure the distance to these specific regions of the LMC and to study the galaxy internal structure.  
%\begin{figure}
%  \includegraphics[width=8cm]{map1211.ps}
%   \caption{The distribution of Classical Cepheids (black dots), HEBs (red dots) and contact binaries (blue dots) in the LMC} 
%  \label{map}
%\end{figure}

\section*{Acknowledgments}
We are grateful to Prof. Slavek Rucinski for carefully reading the manuscript and for his valuable comments.
This work was partially supported by the Gaia Research for
European Astronomy Training (GREAT-ITN) Marie Curie network, funded
through the European Union Seventh Framework Programme ([FP7/2007-2013]
under grant agreement n. 264895.
We warmly thank P. Montegriffo for the development and maintenance of the GRATIS software.
This work made use of EROS-2 data, which were kindly provided by the EROS 
collaboration. The EROS (Exp\'erience pour la Recherche d'Objets Sombres) 
project was funded by the CEA and the IN2P3 and INSU CNRS institutes.
We acknowledge the OGLE team for making public their catalogues.
We thank the UK VDFS team 
% the Cambridge Astronomy Survey Unit (CASU) and the VISTA Science Archive at Wide Field Astronomy Unit (Edinburgh) (WFAU) for providing calibrated data products under the support of the STFC. %Maria-Rosa's comment
for providing calibrated data products under the support of the STFC.
Financial support for this work was provided by PRIN-INAF 2008 (P.I. M. Marconi), COFIS ASI- INAF I/016/07/0 (P.I. M. Tosi) and by PRIN-MIUR 2010 (2010LY5N2T) ``Chemical and Dynamical evolution of the Milky Way and Local Group galaxies" (PI F. Matteucci). 
RdG acknowledges research support from the National Natural Science Foundation of China (NSFC) through grant 11373010. This work was partially supported by the Argentinian institutions CONICET and Agencia Nacional
de Promoci\'on Cient\'{\i}fica y Tecnol\'ogica (ANPCyT).

\section*{Appendix}

Main properties and Fourier parameters of the light curves for 493 new HEBs and for 1275 HEBs which were already detected by other surveys are presented in Tables~\ref{tab_gen_n} and \ref{tab_gen_o}, respectively. The tables provide the EROS-2 identification numbers (column 1) and coordinates (RA and DEC at J2000; columns 2 and 3) of the EBs. Periods (column 4) for the majority of stars are from the EROS-2 catalogue, while for  225 sources marked by an asterisk periods were recalculated in this study. Number of digits in the periods are the same as originally listed in the EROS-2 and OGLE~III (for Table~\ref{flames}) catalogues. 
 Mean$\langle B_{EROS}\rangle$ and $\langle R_{EROS}\rangle$ magnitudes are listed in columns 5 and 6 respectively. The EROS-2 team provided us values with three digits as computed  using all observations involved in the period determination (e.g. after excluding outliers), however we rounded them to two digits to account for the typical errors of the individual data-points which vary from 0.02 to 0.08 mag 
 depending on magnitude (see Table~\ref{EROS_phot}). Column 7 lists the epochs of minimum light in the $R_{EROS}$ passband we calculated in this study, they are listed with four digits, 
 in agreement with the actual precision of EROS-2 HJDs (see below).
   Columns from 8 to 13 %of  Tables~\ref{tab_gen_n} and \ref{tab_gen_o} 
  present the parameters of the Fourier decomposition  in the $R_{EROS}$ passband calculated in this study. %HJDs provided by the EROS-2 catalogue are accurate to within 10 {\it s} hence epochs of minimum light have four digit accuracy .
We provide complete time-series data in the EROS-2 passbands of all 1768 EBs from the EROS-2 catalogue in the electronic version of the paper. 
As an example, Table~\ref{EROS_phot} presents  the  $R_{EROS}$ and $B_{EROS}$  time-series photometry and related errors for  the eclipsing binary lm0023n11843. All magnitude values  are listed with 3 digits as originally given to us  by the EROS-2 team. Similarly, HJDs are listed with 5 digits as provided by the EROS-2 catalogue, however they are accurate to within 10 {\it s}. 
Complete time-series data in the $K_{\rm s}$-band for 999 EBs in common between the EROS-2 and VMC catalogues is available in the electronic version of the paper.
As an example, Table~\ref{VMC_phot} shows  the $K_{\rm s}$-band time series data and related errors for EB lm0023n11843. All magnitude values  are listed with 3 digits as originally listed in  the VSA archive.%is presented in Table~\ref{VMC_phot}. 
 HJDs of the $K_{\rm s}$ time-series data are accurate to within 6 digits.
In the electronic version of the tables missing data are always marked as 99.999 (for magnitudes) and 9.999 (for errors). 

Finally, Table~\ref{tab_K} provides information about the cross-identifications (EROS-2 and VMC IDs)  for 999 EBs in common between the two catalogues, their periods and the  $K_{\rm s}$ and $R_{EROS}$ magnitudes at maximum light.
\newpage 

\begin{table*}
 %\centering
 \begin{minipage}{200mm}
 \caption{Identification, main properties and Fourier parameters of the light curves for the 493 new HEBs in the LMC.
%Column~1: EROS-2 id of the star
%Column~2: RA from the EROS-2 catalogue;
%Column~3: DEC from the EROS-2 catalogue;
%Column~4: Period ($^{a}$- Stars for which a new period estimate is given. See text for the details.);
%Column~5: Mean magnitude in the $R_{EROS}$ passband;
%Column~6: Mean magnitude in the $B_{EROS}$ passband;
%Column~7: Colour $\langle B_{EROS}\rangle$-$\langle R_{EROS}\rangle$;
%Column~8: Epoch of the primary minimum;
%Column~9-Column~14: Fourier parameters of the light curves.
}
 \label{tab_gen_n}

  \tiny
   \begin{tabular}{@{}l|l|l|c|c|c|c|c|l|l|l|l|l|l|@{}}
\hline
~~EROS-2 id &   ~~~~RA   & ~~~~DEC  & Period& $\langle R_{EROS}\rangle$  & $\langle B_{EROS}\rangle$ &Epoch(min)& $a_0$&$~~~~a_1$&~~~~$a_2$&~~~~$a_3$&~~~~$a_4$&~~~~$a_5$\\
 {} & ~~~~(J2000)& ~~~~(J2000)  & {}   &  {}& {}   & (HJD$-$2,450,000)& {}&{}&{}&{}&{}&{} \\ 
  {} & ~~~~(deg)& ~~~~(deg)  & (day)     &  (mag)  & (mag)   & {}& {}&{}&{}&{}&{}&{} \\ 
 \hline
lm0323n20546   &80.4206    &$-$66.8744     &0.900708    &17.24  & 17.06    &1851.6263	 &0.879	&$-$0.028    &$-$0.167     &$-$0.014      &$-$0.062	&$-$0.0030 \\    
lm0341k8979      &83.66655 &$-$66.30307   &0.911721    &17.00  & 16.75    &1173.7244	 &0.882	&$-$0.032    &$-$0.173      &$-$0.011     &$-$0.074	&~~0.0	    \\
lm0344m26510  &83.0365    &$-$67.12818   &0.912445    &17.32  & 17.18    &1751.8628	 &0.917	&$-$0.01       &$-$0.117     &~~0.0020         &$-$0.052	&~~0.0060      \\
lm0342k18196   &82.89531  &$-$66.75002   &0.941447    &17.03  & 16.82    &~434.8095  &0.911	&$-$0.0060   &$-$0.111    &$-$0.0010   &$-$0.045	&$-$0.0050 \\    
lm0346l14981    &82.79685  &$-$67.5609     &0.941548    &17.04  & 17.07    &1657.5184	 &0.985	&$-$0.025     &$-$0.035    &$-$0.011     &$-$0.017	&$-$0.0	\\    
lm0341k18581   &83.73331  &$-$66.3684     &0.947627    &17.37  & 17.22    &1981.6242	 &0.985	&$-$0.033     &$-$0.0060  &$-$0.0050   &$-$0.0050	 &~~0.0030 \\     
lm0341k4660     &83.61982  &$-$66.27231   &0.953374    &17.34  & 17.09    &1869.6179	 &0.932	&$-$0.0070   &$-$0.107    &$-$0.011     &$-$0.069	&$-$0.0050\\     
lm0340m15852  &83.06971 &$-$66.35575   &0.958465    &17.46  & 17.28    &2310.7127	 &0.979	&$-$0.062     &$-$0.019    &$-$0.017     &$-$0.0060	&~~0.0070     \\ 
lm0542k20454   &72.90434  &$-$70.96458   &0.965838    &17.37  & 17.32    &1659.4784	 &0.975	&$-$0.048     &$-$0.039    &$-$0.032     &$-$0.02	&$-$0.0040 \\    
 lm0362n18001$^{*}$   &86.56198 &  $-$66.88331  &1.950688   & 17.32  & 17.31  &  1125.6826 &0.978   &$-$0.0040   &$-$0.023      &$-$0.0020   &$-$0.0010  & $-$0.0   \\
\hline
\end{tabular}
\end{minipage}
\medskip

$^{*}$ Star with period re-determined based on our study of the light curve.~~~~~~~~~~~~~~~~~~~~~~~~~~~~~~~~~~~~~~~~~~~~~~~~~~~~~~~~~~~~~~~~~~~~~~~~~~~~~~~~~~~~~~~~~~\\
The table is ~published ~in ~its ~entirety as ~Supporting ~Information ~with the ~electronic ~version of the ~article. A portion is shown here for guidance regarding its form and content.~~~~~~~~~~~~~~~~~~~~~~~~~~~~~~~~~~~~~~~~~~~~~~~~~~~~~~~~~~~~~~~~~~~~~~~~~~~~~~~~~~~~~~~~~~~~~~~~~~~~~~~~~~~~~~~~~~~~~~~~~~~~~~~~~~~~~~~
\
\end{table*}

%\clearpage

\begin{table*}
% \centering
 \begin{minipage}{180mm}
 \caption{Identification, main properties and Fourier parameters of the light curve for  the 1275 HEBs in the LMC  which were already detected by other surveys.
%Column~1: EROS-2 id of the star
%Column~2: RA from the EROS-2 catalogue;
%Column~3: DEC from the EROS-2 catalogue;
%Column~4: Period ($^{a}$- Stars for which a new period estimate is given. See text for the details.);
%Column~5: Mean magnitude in the $R_{EROS}$ passband;
%Column~6: Mean magnitude in the $B_{EROS}$ passband;
%Column~7: Colour $\langle B_{EROS}\rangle$-$\langle R_{EROS}\rangle$;
%Column~8: Epoch of the primary minimum;
%Column~9-Column~14: Fourier parameters of the light curves.
}
 \label{tab_gen_o}

  \tiny
   \begin{tabular}{@{}l|l|l|l|c|c|c|c|c|c|l|c|c|@{}}
\hline

~~EROS-2 id &   ~~~~RA   & ~~~~DEC  & Period& $\langle R_{EROS}\rangle$  & $\langle B_{EROS}\rangle$ &Epoch(min)& $a_0$&$~~~~a_1$&~~~~$a_2$&~~~~$a_3$&~~~~$a_4$&~~~~$a_5$\\
 {} & ~~~~(J2000)& ~~~~(J2000)  & {}   &  {}& {}   & (HJD$-$2,450,000)& {}&{}&{}&{}&{}&{} \\ 
  {} & ~~~~(deg)& ~~~~(deg)  & (day)     &  (mag)  & (mag)   & {}& {}&{}&{}&{}&{}&{} \\ 

 \hline
lm0555k12721$^{*}$ &76.2287~   &$-$71.23919    &0.8010095   &17.26	&17.08	    &2184.7009	   &0.957   &$-$0.037    &$-$0.055    &$-$0.0070   &$-$0.016   &~~0.0030   \\
lm0354n8770               &85.11371   &$-$67.16454    &0.904003     &17.64	&17.40       &1185.7298	   &0.913   &$-$0.03~    &$-$0.127    &$-$0.0080   &$-$0.041    &~~0.0020   \\
lm0193k19182             &79.73758   &$-$68.10716    &0.905358     &17.05	&16.99	    &2213.7821	   &0.958   &$-$0.01~    &$-$0.047    &$-$0.0060   &$-$0.024   &~~0.0060   \\
lm0285n10229             &73.41711   &$-$67.17722    &0.911792     &17.07	&16.85       &1532.5773	   &0.973   &$-$0.015    &$-$0.02~    &$-$0.0010   &~~~~0.0010  &$-$0.0020  \\
lm0023n11843             &83.3321~   &$-$69.62209    &0.912568     &17.46	&17.39       &1701.4807	   &0.854   &$-$0.033    &$-$0.141    &$-$0.01        &$-$0.032    &~~0.017~    \\
lm0123m11836            &73.49311   &$-$69.46511    &0.919332     &17.18	&17.29	    &2519.7319	   &0.97~   &$-$0.014    &$-$0.037    &$-$0.0050   &$-$0.013   &$-$0.0010  \\
lm0122n13303             &72.46236   &$-$69.64522    &0.922604     &17.17	&16.94	    & ~438.8273	   &0.947   &$-$0.023    &$-$0.08~    &$-$0.01       &$-$0.018   &~~0.0040   \\
lm0030n21391             &84.32319   &$-$69.38179    &0.922672     &16.88	&16.77       &2304.7686	   &0.883   &$-$0.043    &$-$0.153   &$-$0.018      &$-$0.046    &$-$0.0010  \\ 
lm0226n20767             &84.60396   &$-$69.00125    &0.923941     &17.65	&17.45       &1975.6799	   &0.913   &$-$0.013    &$-$0.12~   &$-$0.0050   &$-$0.056    &$-$0.0~~~~ \\  
lm0106n13876             &76.54384   &$-$70.34353    &0.925344     &17.51	&17.36       &~498.6062	   &0.892   &$-$0.033    &$-$0.133   &$-$0.02        &$-$0.043    &$-$0.0010\\  
\hline
\end{tabular}
\end{minipage}
\medskip

$^{*}$ Star with period re-determined based on our study of the light curve.~~~~~~~~~~~~~~~~~~~~~~~~~~~~~~~~~~~~~~~~~~~~~~~~~~~~~~~~~~~~~~~~~~~~~~~~~~~~~~~~~~~~~~~~~~\\
The table is ~published ~in ~its ~entirety as ~Supporting ~Information ~with the ~electronic ~version of the ~article. A portion is shown here for guidance regarding its form and content.~~~~~~~~~~~~~~~~~~~~~~~~~~~~~~~~~~~~~~~~~~~~~~~~~~~~~~~~~~~~~~~~~~~~~~~~~~~~~~~~~~~~~~~~~~~~~~~~~~~~~~~~~~~~~~~~~~~~~~~~~~~~~~~~~~~~~~~
\

\end{table*}

\begin{table*}

\begin{center}

  \caption{EROS-2 HEBs with  existing optical spectroscopy; E06 \citep{Ev2006}; W14 \citep{Wal2014}; E14 (Evans et al. in prep). OGLE~III periods are from \citet{Grac2011}. SB1 and SB2 stand for single- and double-lined spectroscopic binaries, respectively.}\label{flames}

\begin{tabular}{lllccllll}

\hline\hline
EROS-2 id & RA(J2000)    & DEC(J2000)   &  \multicolumn{2}{c}{Period (day)} &  Alternative id & Spectral type & Notes & Ref. \\ 
         & (deg) & (deg) &  EROS-2     & OGLE~III          &            &               &       & \\
\hline

lm0290l18998 &73.88709  & $-$66.54208 & 3.224805 & $-$      & N11-107       & B1-2\,$+$\,Early B       & SB2    & E06 \\
lm0290l5213  &73.95604 & $-$66.43437 & 1.791025 & $-$      & N11-119       & B1.5 V                   & SB2    & E06 \\
lm0344l12773 & 82.6699  & $-$67.19545 & 4.952487 & $-$      & NGC\,2004-079 & B2 III                   & SB1    & E06 \\
lm034l21656  & 82.78869 & $-$67.25619  & 4.164156 & $-$      & NGC\,2004-094 & B2.5 III                 & Binary & E06 \\
lm0030m4163  & 84.37804 & $-$69.08817 &2.333416 & 2.333427 & VFTS\,061     & ON8.5III:\,$+$\,O9.7: V: & SB2    & W14 \\
lm0030m3468  & 84.42029 & $-$69.07812 & 1.674098 & 1.674119 & VFTS\,112     & Early B\,$+$\,Early B    & SB2    & E14 \\
lm0030m9744  & 84.46997 & $-$69.16274  & 1.434738 & 1.434745 & VFTS\,189     & B0.7: V                  & Binary & E14 \\
lm0226n24168 & 84.66296 & $-$69.02808 & 1.176008 & 1.176008 & VFTS\,462     & B0.5-0.7 V               & $-$ & E14 \\
lm0426m23482 &75.07795 & $-$66.06284 &  2.345573& $-$	&$-$ & B1: V	  & SB2? & \ldots\\
lm0294m4825  &74.53203 &$-$66.98277 &  2.97779 & 2.9778	& $-$ & B0-0.5 V  & SB? &\ldots\\
lm0436l19007 &76.28093 &$-$66.19386 &  3.301123& $-$	&$-$ & B2 V	  & $-$	&\ldots\\
lm0020n19615 &82.67397 &$-$69.32445 &  4.585353& 4.585031 &$-$   & B1.5 Ib   & SB1?& \ldots\\
lm0031l22987 &85.19681 &$-$69.34126 &  5.413977& 5.414011 &$-$  & B1 III    & SB2 &\ldots\\

\hline

\end{tabular}
\end{center}

\end{table*}

\clearpage

\begin{table}
%\centering
% \begin{minipage}{200mm}
 \caption{Sample $R_{EROS}$ and $B_{EROS}$  time-series photometry for EROS-2   
 EB lm0023n11843 (VMC J053319.72-693719.70).} 
 \label{EROS_phot}
 \begin{tabular}{@{}c|c|c|c|c|@{}}
\hline
HJD $-$ 2,450,000     &  $R_{EROS}$ & Err$_{R_{EROS}}$& $B_{EROS}$ & Err$_{B_{EROS}}$\\
(day) & (mag)& (mag)& (mag)& (mag)\\
\hline
 & & & & \\
 
303.90001 & $-$        & $-$   & 17.276 & 0.041  \\ 
315.89299 & 17.294 & 0.046& 17.012 & 0.038 \\ 
325.82966 &17.431  & 0.050& $-$        & $-$     \\
327.82295 &17.219  & 0.030& 16.973 & 0.029 \\ 
334.87162 &17.546  & 0.029& 17.261 & 0.025 \\ 
351.82713 &17.571  & 0.041& $-$        & $-$     \\ 
361.85427 &17.606  & 0.063& 17.448 & 0.047 \\ 
367.69923 & $-$        & $-$   & 17.168 & 0.041  \\ 
373.86849 &17.234  & 0.033& 16.977 & 0.027 \\ 
377.77906 &17.537  & 0.030& 17.280 & 0.024 \\ 
\hline
\end{tabular}
%\end{minipage}
\medskip

The table is published in its entirety as Supporting Information with the electronic version of the article. A portion is shown here for guidance regarding its format and content.
\end{table}

\begin{table}
 \centering
% \begin{minipage}{200mm}
 \caption{Sample $K_{\rm s}$  time-series photometry for EROS-2  
 EB  lm0023n11843 (VMC J053319.72-693719.70).} 
 \label{VMC_phot}
 \begin{tabular}{@{}lc|c|c|@{}}
\hline

HJD $-$ 2,400,000 & $K_{\rm s}$ & Err$K_{\rm s}$\\
~~~~~~~~(day) & (mag)& (mag)\\
\hline
& & \\
55140.763187   &  17.406   &   0.055 \\  
55143.762677   &  17.435   &   0.047 \\
55147.802823   &  17.573   &   0.052 \\ 
55152.818276   &  17.607   &   0.065 \\ 
55155.725147   &  17.314   &   0.054 \\ 
55161.842037   &  17.677   &   0.076 \\ 
55164.789981   &  17.333   &   0.042 \\ 
55172.764230   &  17.575   &   0.047 \\ 
55191.753090   &  17.336   &   0.041 \\ 
55209.682223   &  17.412   &   0.049 \\ 
\hline
\end{tabular}
%\end{minipage}
\medskip

The table ~is published in ~its entirety as Supporting Information with the electronic version of the article. A portion is shown here for guidance regarding its form and content.~~~~~~~~~~~~~~~~~~~~~~~~~~~~
\end{table}

\begin{table*}
%\centering
% \begin{minipage}{100mm}
 \caption{$R_{\rm EROS}$ and $K_ {\rm s}$ magnitudes at maximum light of EB stars in the LMC from the EROS-2 and VMC data.}
 \label{tab_K}
 \begin{tabular}{@{}ll|c|c|c|c|@{}}
\hline
~~EROS-2 id &~~~~~~~~VMC id & Period & $K_ {\rm s, max}$ &  $R_{\rm EROS, max}$\\
{}&{}& (day) &(mag) &(mag)\\
\hline
  lm0323n20546               &VMC J052140.98-665227.93     &0.900708  &17.116   &17.057\\		
  lm0344m26510              &VMC J053208.76-670741.80     &0.912445  &17.260   &17.161\\	     
  lm0023n11843               &VMC J053319.72-693719.70     &0.912568  &17.314   &17.143\\	     
  lm0030n21391               &VMC J053717.63-692254.37     &0.922672  &16.754   &16.644\\	     
  lm0226n20767               &VMC J053825.02-690004.55     &0.923941  &17.497   &17.560\\	     
  lm0231k9063                 &VMC J054830.84-674223.25     &0.927999  &16.992   &16.883\\	     
  lm0214n10459               &VMC J053127.29-683413.65     &0.938971  &17.652   &17.635\\	     
  lm0171m16733$^{*}$  &VMC J050444.08-674447.89     &0.939001  &15.982   &15.748\\	     
  lm0127k12134               &VMC J045304.25-701051.19     &0.939454  &16.700   &17.384\\	     
  lm0342k18196               &VMC J053134.88-664500.10     &0.941447  &17.195   &16.915\\	     
\hline
\end{tabular}
%\end{minipage}
\medskip

$^{*}$ Star with period re-determined based on our study of the light curve.\\
The table is published in its entirety as Supporting Information with the electronic version of the article. A portion is shown here for guidance regarding its format and content.
\end{table*}

\label{lastpage}

\end{document}